\documentclass[reprint]{revtex4-2}
\usepackage[utf8]{inputenc}
\usepackage{hyperref}
\usepackage{xcolor}
\usepackage{caption}
\usepackage{subcaption}

\usepackage{tikz}
\usetikzlibrary{quantikz}

\usepackage{bbold}

\begin{document}

\title{Compilation of a simple chemistry application to quantum error correction primitives}
\author{Nick S. Blunt}
\author{Gy\"{o}rgy P. Geh\'{e}r}
\author{Alexandra E. Moylett}
\email{alex.moylett@riverlane.com}
\affiliation{Riverlane, St Andrews House, 59 St Andrews Street, Cambridge, CB2 3BZ, UK}
\date{\today}

\begin{abstract}
A number of exciting recent results have been seen in the field of quantum error correction. These include initial demonstrations of error correction on current quantum hardware, and resource estimates which improve understanding of the requirements to run large-scale quantum algorithms for real-world applications. In this work, we bridge the gap between these two developments by performing careful estimation of the resources required to fault-tolerantly perform quantum phase estimation (QPE) on a minimal chemical example. Specifically, we describe a detailed compilation of the QPE circuit to lattice surgery operations for the rotated surface code, for a hydrogen molecule in a minimal basis set. We describe a number of optimisations at both the algorithmic and error correction levels. We find that implementing even a simple chemistry circuit requires 1,000 qubits and 2,300 quantum error correction rounds, emphasising the need for improved error correction techniques specifically targeting the early fault-tolerant regime.
\end{abstract}

\maketitle

\section{Introduction}

Quantum error correction (QEC), the study of how many noisy physical qubits are used to represent a smaller number of less noisy logical qubits, has seen significant recent developments in a number of directions. One such success is experimental demonstrations of error correction successfully suppressing errors on a real-world quantum device \cite{Acharya2023GoogleExperiment}. Another recent development is in careful resource estimates, which have allowed for more accurate estimates of the resources a quantum computer requires to solve problems of significant interest, from estimating chemical properties \cite{Reiher2017, Lee2021TensorHypercontraction, Blunt2022, Kim2022LiIonBatteries, Ivanov2023SolidState} to factoring RSA integers \cite{Gidney2021, litinski2023EllipticCurve}. These developments together have helped define both the current state of our abilities to suppress noise on quantum devices, and where we need to get to in order to solve key industrial problems.

There are some natural next steps following the experimental demonstration of a logical quantum memory. Natural follow-ups include implementing basic logical gates: implementing Pauli gates through transversal operations, non-Pauli Clifford gates through lattice surgery techniques \cite{Horsman2012, Fowler2018, Litinski2018, Litinski2019, Chamberland2022, Gidney2023, Watkins2023LatticeSurgeryCompiler}, and non-Clifford gates initially through error mitigation techniques \cite{Piveteau2021MitigatedT} and later through magic state distillation \cite{Bravyi2005, Gidney2019, Litinski2019magicstate}. Eventually, a natural goal will be to demonstrate small-scale quantum algorithms, showing that these logical operations can be used to solve a toy application. Understanding the resources required for such an algorithm is important for knowing the point at which small applications can start being solved on fault-tolerant quantum computers, as well as helping us understand the constant factors in the scaling of large quantum algorithms. A number of algorithms have recently been proposed that are aimed specifically at this regime, referred to as ``early fault-tolerant'' algorithms \cite{Campbell2022, Lin2022, Wang2022, Ding2022, Wang2023}; it is therefore particularly relevant to assess how challenging even minimal applications will be to perform using fault-toleration operations.

\begin{figure*}
    \begin{subfigure}[b]{0.45\linewidth}
        \includegraphics[width=\linewidth]{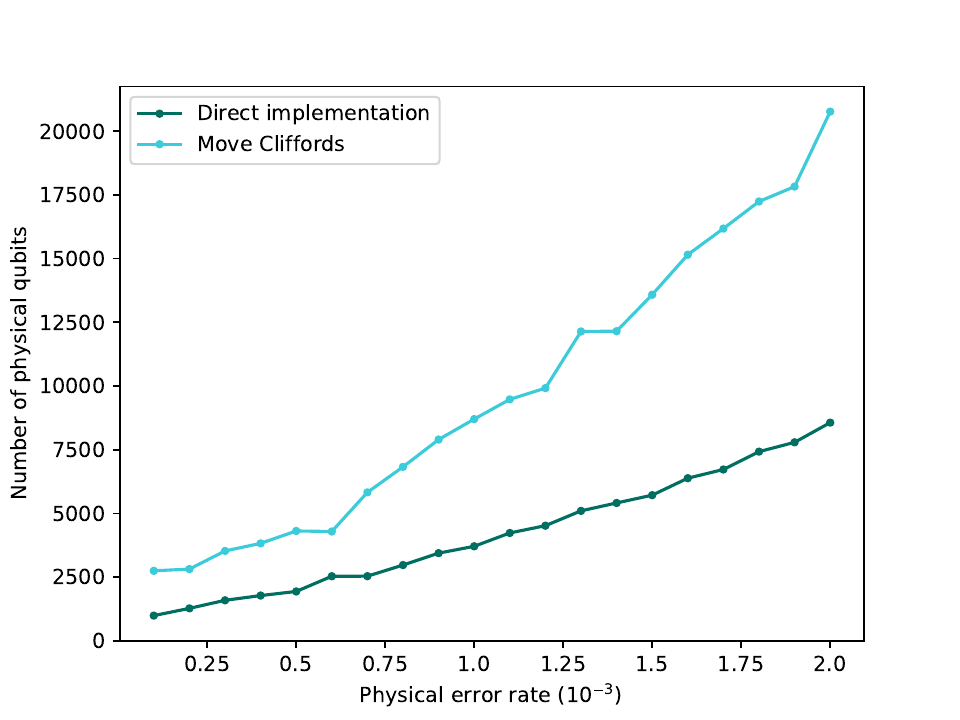}
    \end{subfigure}
    \begin{subfigure}[b]{0.45\linewidth}
        \includegraphics[width=\linewidth]{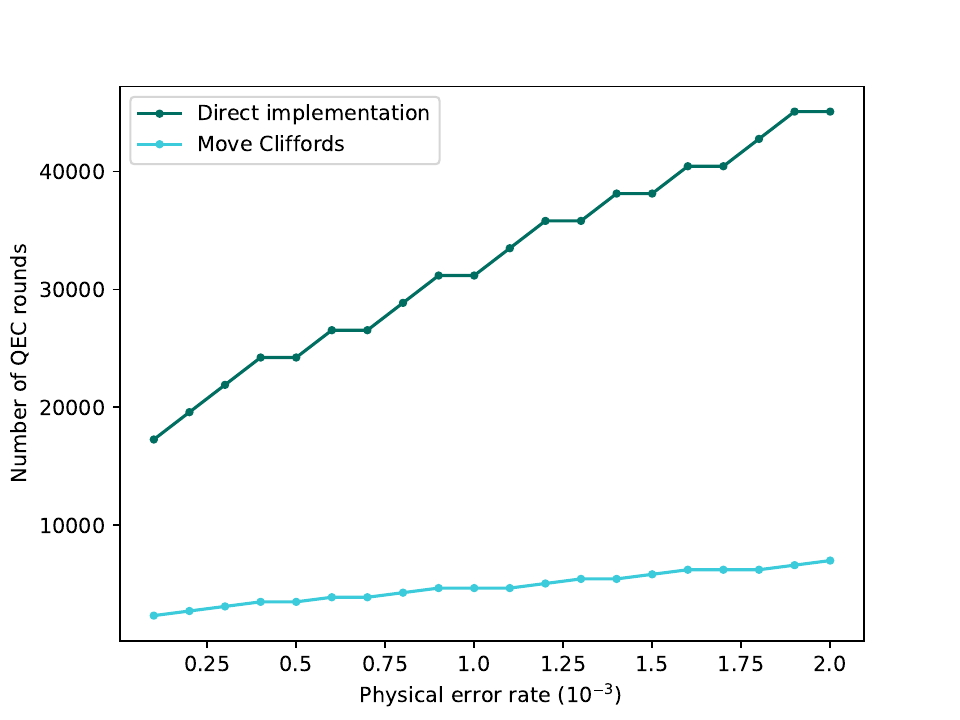}
    \end{subfigure}
    \caption{Estimated cost in physical qubits and time for calculating the ground-state energy of a hydrogen molecule on an error-corrected quantum computer with varying physical error rates, using iterative quantum phase estimation. Methods for implementing logical gates through either directly implementing Clifford and $T$ gates or moving Clifford gates through the circuit are described in Sections \ref{ssec:direct-impl-clifford} and \ref{ssec:commute-clifford}, respectively. Further details about estimating these resource requirements are presented in Section \ref{ssec:results}.}
    \label{fig:results-summary}
\end{figure*}

In this work, we estimate the resources required for implementing a small quantum algorithm on a fault tolerant quantum computer, including detailed consideration of how to perform each required operation using lattice surgery. The application we choose is quantum phase estimation (QPE) applied to finding the ground-state energy of the hydrogen molecule. This application is sufficiently small that related circuits without QEC have already been successfully run on current quantum hardware \cite{Graham2022, Blunt2023Rigetti}. We investigate optimisations of this algorithm at a variety of levels, including algorithmic \cite{Wecker2015,Reiher2017, Kivlichan2020,McArdle2022}, gate decompositions \cite{Ross2016}, compilation to lattice surgery primitives \cite{Horsman2012, Fowler2018, Litinski2018, Litinski2019, Chamberland2022, Gidney2023} and generation of magic states \cite{Litinski2019magicstate}. Our final resource estimates are presented in Fig.\ \ref{fig:results-summary}, looking at different physical error rates and techniques which trade off time and space resource requirements. It is worth noting that when implemented on the surface code, even this small application requires hundreds of physical qubits and thousands of QEC rounds. This shows the significant prefactor associated with quantum error correction, and suggests that in early fault-tolerance further techniques will be required to yield small-scale algorithmic demonstrations \cite{Piveteau2021MitigatedT}.

The rest of this paper is laid out as follows. In Section \ref{sec:logical-circuit}, we review the algorithms and chemical system to be considered, and present the logical quantum circuit. In Section \ref{sec:implementing-logical-gates}, we describe how to decompose the logical quantum circuits into operations from the Clifford+$T$ gate set and how to implement these gates on the surface code using lattice surgery primitives. In Section \ref{sec:qec-overheads} we estimate the overhead introduced by quantum error correction. Finally, we conclude with some open questions and further directions for research in Section \ref{sec:conclusion}.

\section{Logical quantum circuit}
\label{sec:logical-circuit}

\subsection{Quantum phase estimation}

We begin with a brief introduction to the quantum algorithms considered in this study, which are two types of quantum phase estimation (QPE)  \cite{kitaev_quantum_1995}. QPE is one of the key proposed quantum algorithms for calculating ground and excited-state energies in electronic structure problems. Provided an initial trial state can be prepared that has a sufficiently good overlap with the true ground state (which is usually the case for molecular systems), QPE is capable of obtaining energy estimates to a desired precision in polynomial time with system size. However, the algorithm requires high circuit depths for non-trivial examples, and so has seen less attention compared to variational quantum algorithms in current NISQ applications. For fault-tolerant applications, however, it is often regarded as the algorithm of choice.

\begin{figure}
\begin{subfigure}[b]{\linewidth}
\includegraphics[width=\linewidth]{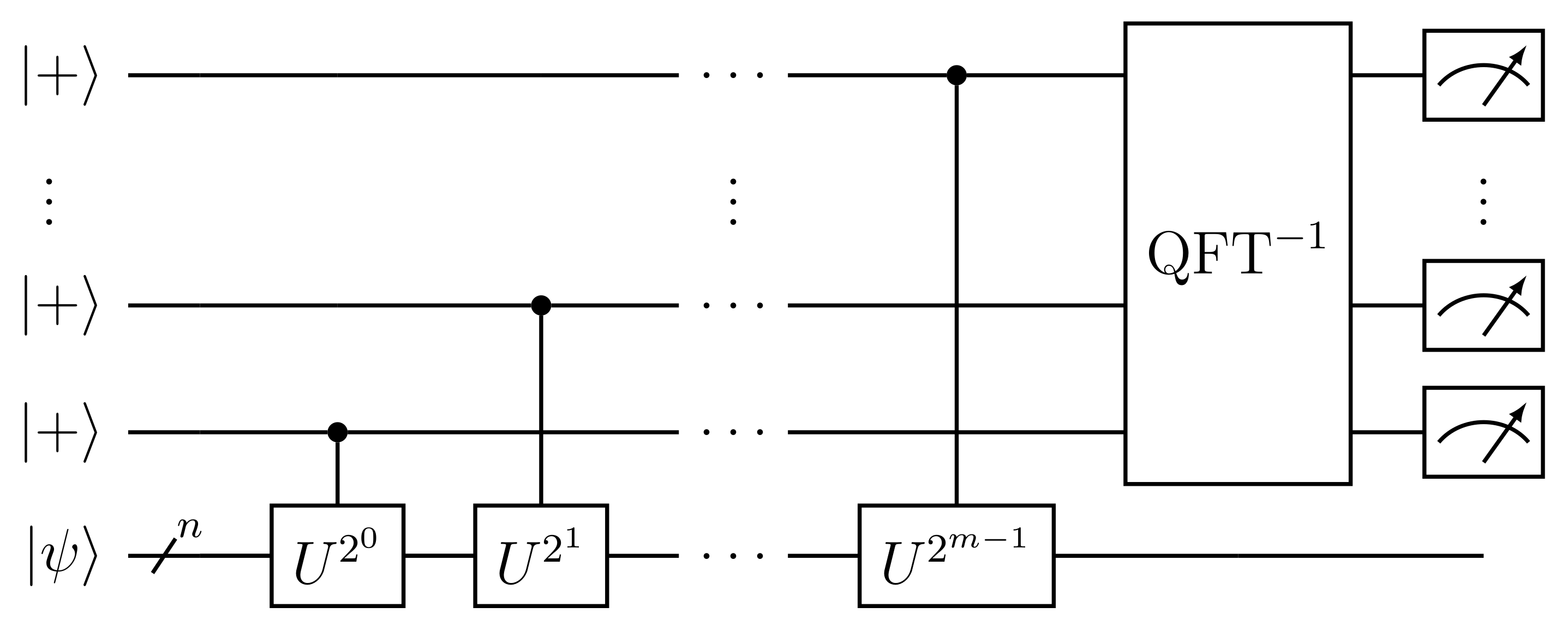}
\caption{Textbook QPE circuit.}
\label{fig:textbook_qpe_circuit}
\end{subfigure}

\par\bigskip

\begin{subfigure}[b]{\linewidth}
\begin{center}
\begin{quantikz}[row sep=0.15cm]
\lstick{$\ket{+}$} & \qw & \ctrl{1} & \gate{R_z(\omega_k)} & \gate{H} & \meter{} & \rstick{\scriptsize k'th bit} \\
\lstick{$\ket{\psi}$} & \qw\qwbundle{n} & \gate[wires=1][0.8cm][0.7cm]{U^{2^{k-1}}} \qw & \qw & \qw & \qw
\end{quantikz}
\end{center}
\caption{Iterative QPE circuit.}
\label{fig:iqpe_circuit}
\end{subfigure}
\caption{QPE circuits used in this paper. In both cases, the state $|\psi\rangle$ is over $n$ qubits. In (b), the circuit is iterated backwards from $k=m$ (in the initial iteration) to $k=1$ (in the final iteration). The rotation angle in iteration $k$ is $\omega_{k} = -\pi(0.x_{k+1}x_{k+2}\ldots x_{m})$, with $\omega_m = 0$ in the initial iteration. While the ancilla is measured at the end of each iteration, the data qubits remain coherent throughout.}
\label{fig:qpe_circuit}
\end{figure}

We focus on the ``textbook'' \cite{Nielsen2010} and iterative (semi-classical) QPE algorithms \cite{cleve_1998, mosca_1999, childs_2000, iqpe_2007}. The textbook QPE algorithm is perhaps the best known QPE approach, the circuit for which is presented in Fig.~\ref{fig:textbook_qpe_circuit}. The algorithm allows one to measure the eigenphases of some unitary $U$ up to $m$ bits of precision; doing so requires $m$ ancilla qubits, in addition to the $n$ data qubits needed to represent $U$. At the end of the circuit, an inverse quantum Fourier transform (QFT) is performed and the ancilla qubits are measured. If the input state $|\psi\rangle$ is an exact eigenstate of $U$, then the measured bits will yield the bits of the corresponding eigenphase. For a non-exact $|\psi\rangle$, the probability of obtaining the desired phase will depend on the overlap between $|\psi\rangle$ and the corresponding exact eigenstate.

The inverse QFT can also be performed in a semi-classical manner \cite{Griffiths1996}. Using such a semi-classical QFT, the resulting phase estimation algorithm is performed iteratively, obtaining one bit of information about the phase from each iteration. We refer to this approach as iterative quantum phase estimation \cite{iqpe_2007}. Iterative QPE has many of the benefits of the textbook approach, including a Heisenberg-limited running time $\mathcal{O}(\epsilon^{-1})$ for a precision of $\epsilon$, but has the significant benefit that it uses only a single ancilla qubit.

We briefly give some analysis of the iterative QPE approach here. We are interested in estimating the eigenvalues of a Hamiltonian
\begin{equation}
    H = \sum_{j=1}^L c_j P_j,
    \label{eq:hamil_1}
\end{equation}
where $P_j$ are $n$-qubit Pauli operators and $c_j$ are coefficients. We denote the eigenvalues and eigenvectors of $H$ by $\{\lambda_j ; |\Psi_j\rangle \}$. We multiply $H$ by a constant $t$ such that $-0.5 \le \lambda_j t \le 0.5$ for all $j$, which can always be achieved by choosing $ 1 / t = 2\sum_j |c_j|$. We then work with the unitary
\begin{equation}
    U = e^{2 \pi i \, H t}.
\end{equation}
The eigenvalues of $U$ are $e^{2 \pi i \phi_j}$, where the range $0 \le \phi_j \le 1$ can be chosen. It is then simple to obtain $\lambda_j t$ from $\phi_j$, which only differ due to the wrapping of phases; the normalization of $H t$ above is chosen to avoid potential ambiguity in this wrapping. Therefore each $\phi_j$ can be written in binary as
\begin{equation}
    \phi_j = 0.\phi_{j1} \phi_{j2} \ldots \phi_{jm} \ldots.
\end{equation}
In iterative QPE the bits $\phi_{jk}$ are measured directly using the circuit in Fig.~\ref{fig:iqpe_circuit}. The circuit is performed for $m$ iterations in order to obtain $m$ bits of precision for $\phi_j$, starting with $k=m$ and iterating backwards to $k=1$. After each controlled-unitary operation, an $R_z(\omega_k)$ gate is applied to the ancilla with angle
\begin{equation}
    \omega_k = -\pi(0.x_{k+1}x_{k+2}\ldots x_{m}),
\end{equation}
which depends on the measurement results from previous iterations (and $\omega_m = 0$ in the initial iteration).

The data qubits are prepared in an initial state $|\psi\rangle$, which should be an approximation to the exact state whose energy is to be estimated. We write $|\psi\rangle$ in the eigenbasis of $H$ by
\begin{equation}
    |\psi\rangle = \sum_j \nu_j |\Psi_j\rangle.
\end{equation}

The state of the qubits before the first measurement ($k=m$) is then
\begin{equation}
    \frac{1}{2} \sum_j \nu_j \bigg[ (1 + e^{i 2^m \pi \phi_j}) |0\rangle + (1 - e^{i 2^m \pi \phi_j}) |1\rangle \bigg] \otimes |\Psi_j\rangle.
\end{equation}
Consider the simple case where $\phi_j$ can be represented by exactly $m$ bits, so that $\phi_j = 0.\phi_{j1} \phi_{j2} \ldots \phi_{jm} 0 0 \ldots$. In this case $\exp(i 2^m \pi \phi_j) = \exp(i \pi \phi_{jm})$ exactly, and the state of the system before measurement is
\begin{equation}
    \frac{1}{2} \sum_j \nu_j \bigg[ (1 + e^{i \pi \phi_{jm}}) |0\rangle + (1 - e^{i \pi \phi_{jm}}) |1\rangle \bigg] \otimes |\Psi_j\rangle.
\end{equation}
Thus the probabilities of measuring the ancilla as $0$ or $1$ are
\begin{align}
    P_0 &= \sum_j |\nu_j|^2 \mathrm{cos}^2 \Big( \frac{\pi \phi_{jm}}{2} \Big), \\
    P_1 &= \sum_j |\nu_j|^2 \mathrm{sin}^2 \Big( \frac{\pi \phi_{jm}}{2} \Big).
\end{align}
Provided that $|\nu_j|$ is sufficiently large for the desired state $|\Psi_j\rangle$, the desired bit will be measured with high probability. The measurement will also project away the contribution from those states $|\Psi_j\rangle$ for which $\phi_{jm}$ does not match the measured result. It is simple to continue this process for subsequent iterations to $k=1$. After the final iteration, the probability that all of the bits for the desired $\phi_j$ were measured is $|\nu_j|^2$. Therefore, for a sufficiently good initial state, and a sufficient number of repetitions, the ground-state energy can be measured with high probability. Further clear analysis is given in \cite{iqpe_2007}.

In addition to the textbook and iterative QPE methods, there has been recent progress on statistical phase estimation methods \cite{Somma2019, OBrien2019, Lin2022, Wan2022, Ding2022}. Compared to the above approaches, such statistical methods allow shorter circuit depth \cite{Wang2022, Ding2022} and ready combination with error mitigation techniques \cite{Blunt2023Rigetti}, in exchange for performing many circuits. It has been suggested that these methods are particularly appropriate for early-fault tolerant quantum computers. We do not consider such methods here, but note that they would be interesting to investigate further in the context considered here.

\subsection{Hamiltonian simulation via Trotterization}
\label{ssec:trotter}

In this section we briefly discuss first and second-order Trotterization, and present an optimization to the latter.

\begin{figure*}[t]
(a) \hspace{1mm}
\begin{quantikz}
\lstick{$\ket{+}$} & \ctrl{1} & \qw \\
\lstick{$\ket{\psi}$} & \gate[wires=1][1.1cm][0.8cm]{e^{iHt}} & \qw
\end{quantikz}
\hspace{12mm} (c) \hspace{1mm}
\begin{quantikz}
\lstick{$\ket{+}$} & \ctrl{1}\gategroup[2,steps=2,style={dashed,rounded corners,fill=blue!20, inner xsep=2pt},background]{} & \octrl{1} & \ctrl{1}\gategroup[2,steps=2,style={dashed,rounded corners,fill=blue!20, inner xsep=2pt},background]{} & \octrl{1} & \qw & \ldots \\
\lstick{$\ket{\psi}$} & \gate[wires=1][1.1cm][0.8cm]{e^{i H_1 t/4}} & \gate[wires=1][1.1cm][0.8cm]{e^{-i H_1 t/4}} & \gate[wires=1][1.1cm][0.8cm]{e^{i H_2 t/4}} & \gate[wires=1][1.1cm][0.8cm]{e^{-i H_2 t/4}} & \qw & \ldots
\end{quantikz}

\vspace{6mm}

(b) \hspace{1mm}
\begin{quantikz}
\lstick{$\ket{+}$} & \ctrl{1} & \octrl{1} & \qw \\
\lstick{$\ket{\psi}$} & \gate[wires=1][1.1cm][0.8cm]{e^{iHt/2}} & \gate[wires=1][1.1cm][0.8cm]{e^{-iHt/2}} & \qw
\end{quantikz}
\hspace{15mm} (d) \hspace{1mm}
\begin{quantikz}
\lstick{$\ket{+}$}    & \gate[2]{e^{-i \, Z \otimes H_1 t / 4 }} & \gate[2]{e^{-i \, Z \otimes H_2 t / 4 }} & \qw & \ldots \\
\lstick{$\ket{\psi}$} & \qw                                      & \qw                                      & \qw & \ldots
\end{quantikz}
\hspace{20mm}
\caption{Circuit diagrams demonstrating reduction of the controlled time evolution operator in QPE with second-order Trotterization. (\ref{fig:second_order_reduction}a) The time evolution operator controlled on an ancilla, and acting on an initial trial state $|\psi\rangle$, which can be equivalently replaced by (\ref{fig:second_order_reduction}b) in phase estimation circuits. For the second-order Trotter formula, this can be further reduced to the circuit (\ref{fig:second_order_reduction}c). Lastly, each pair of boxed terms can be expressed as a single multi-qubit Pauli rotation (\ref{fig:second_order_reduction}d).}
\label{fig:second_order_reduction}
\end{figure*}
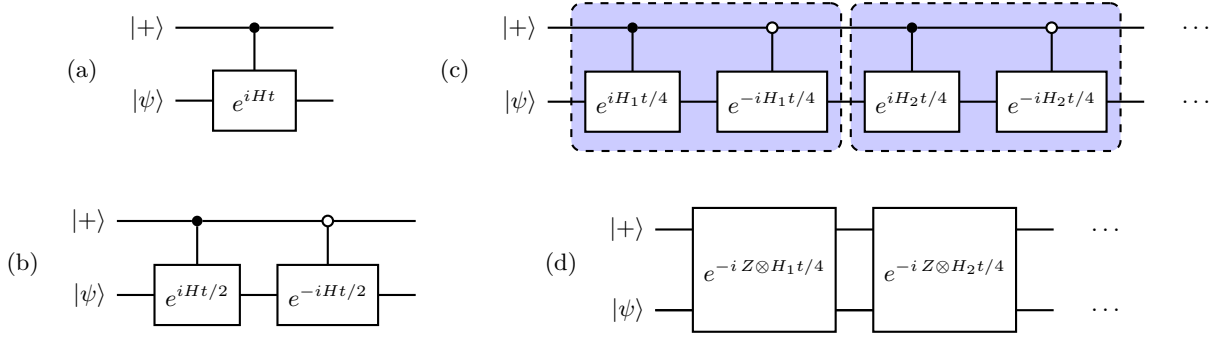

We consider $n$-qubit Hamiltonians of the form of Eq.\ \ref{eq:hamil_1}. We specifically denote $H_j = c_j P_j$, so that
\begin{equation}
    H = \sum_{j=1}^L H_j.
\end{equation}
In Trotter schemes more generally, each $H_j$ might correspond to a sum of commuting Pauli terms, rather than a single Pauli contribution.

We are concerned with implementing an operator $U = e^{iHt}$, controlled on an ancilla qubit. The well-known first- and second-order Trotter approximations, $U_1$ and $U_2$, are
\begin{equation}
    U_1 = \prod_{j=1}^L e^{i H_j t}
\end{equation}
and
\begin{equation}
    U_2 = \prod_{j=1}^L e^{i H_j t/2} \prod_{j=L}^1 e^{i H_j t/2},
\end{equation}
which have errors $\mathcal{O}(t^2)$ and $\mathcal{O}(t^3)$ compared to the exact $U$, respectively.

Let us consider the number of single-qubit rotations needed to implement the controlled $U_1$ and $U_2$ unitaries, as required to perform QPE. Each controlled Pauli rotation, which we shall denote by $W_j$, can be rewritten as
\begin{align}
     W_j &= |0\rangle \langle 0| \otimes \mathbb{1} + |1\rangle \langle 1| \otimes \mathrm{exp}(i \theta_j P_j), \\
     &= \mathrm{exp}(i \theta_j |1\rangle \langle 1 | \otimes P_j), \\
    &= \mathrm{exp}(i (\theta_j/2) (\mathbb{1} - Z) \otimes P_j), \\
        &= \mathrm{exp}(i (\theta_j/2) \mathbb{1} \otimes P_j) \, \mathrm{exp}(-i (\theta_j/2) Z \otimes P_j),
\end{align}
which is a product of two multi-qubit Pauli rotations. These can be reduced to a single-qubit rotation each after conjugation through an appropriate Clifford \cite{Wan2022}. Therefore the cost of each controlled Pauli rotation is $2$ single-qubit rotations plus Cliffords, and the number of single-qubit rotations for $U_1$ is $2L$ per Trotter step.

At first glance it appears that for a given $t$, the second-order formula requires $4L$ single-qubit rotations to implement. In fact in QPE circuits this is not the case, and the second-order formula can also be implemented with $2L$ rotations, as for the first-order formula, but with better error suppression. This trick was introduced in Ref.~\cite{Wecker2015}, and is known as directionally-controlled phase estimation. It was expanded on in Refs.~\cite{Reiher2017, Kivlichan2020} and also used in Ref.~\cite{McArdle2022}.

We briefly give a derivation of the directionally-controlled approach. The general procedure is presented in Fig.~\ref{fig:second_order_reduction}. We consider the controlled time evolution operator in Fig.~\ref{fig:second_order_reduction}(a). The state of the qubits at the end of this circuit is
\begin{equation}
    | \Psi \rangle = \frac{1}{\sqrt{2}} \Big(|0\rangle \otimes | \psi\rangle + |1\rangle \otimes e^{iHt} | \psi \rangle \Big).
\end{equation}
Now, note that we can apply $e^{-iHt/2}$ to the data qubits in Fig.~\ref{fig:second_order_reduction}(a) without affecting any measurement outcomes; since this operator commutes with all controlled-$e^{iHt}$ gates, it can be moved to the end of the circuit where it has no effect on the measurement of the ancilla. With this additional operator applied, the final state of the qubits is
\begin{equation}
    | \Psi \rangle = \frac{1}{\sqrt{2}} \Big( |0\rangle \otimes e^{-iHt/2} | \psi\rangle + |1\rangle \otimes e^{iHt/2} | \psi \rangle \Big),
\end{equation}
and we see that we can work with circuit in Fig.~\ref{fig:second_order_reduction}(b) instead.

We next expand $e^{iHt/2}$ via its Trotter formula,
\begin{equation}
    e^{iHt/2} \approx V_K \ldots V_2 V_1,
\end{equation}
where $K$ is the number of terms in the Trotter product formula, equal to $L$ for the first-order formula and $2L$ for second-order formula. Then,
\begin{align}
    |\Psi\rangle &= |0\rangle \otimes (V_K \ldots V_2 V_1)^{\dagger} |\psi\rangle + |1\rangle \otimes (V_K \ldots V_2 V_1) |\psi\rangle \nonumber \\
    &= |0\rangle \otimes V_1^{\dagger} V_2^{\dagger} \ldots V_K^{\dagger} |\psi\rangle + |1\rangle \otimes V_K \ldots V_2 V_1|\psi\rangle.
\end{align}
For even-order Trotter formulas the string of operators $V_K \ldots V_2 V_1$ is symmetric, so that $V_j=V_{K-j+1}$, and the expansion is unchanged when the order of the terms is reversed. Therefore, for the second-order Trotter formula (but \emph{not} the first-order formula) we can write
\begin{equation}
    |\Psi\rangle = |0\rangle \otimes V_K^{\dagger} \ldots V_2^{\dagger} V_1^{\dagger} |\psi\rangle + |1\rangle \otimes V_K \ldots V_2 V_1|\psi\rangle,
\end{equation}
which is equivalent to the circuit in Fig.~\ref{fig:second_order_reduction}(c). Lastly, note that the paired operators in Fig.~\ref{fig:second_order_reduction}(c) can each be expressed as
\begin{equation}
    e^{i |1\rangle \langle 1| \otimes H_i t/4} e^{-i |0\rangle \langle 0| \otimes H_i t/4} = e^{-i Z \otimes H_i t / 4},
\end{equation}
which can be reduced to a rotation on a single qubit plus Clifford gates. Therefore, application of the second-order Trotter formula in QPE can be performed with $2L$ rotations, which is equal to the number required for the first-order Trotter formula. In addition, the Trotter expansion is applied to the operator $e^{iHt/2}$ instead of $e^{iHt}$, resulting in lower Trotter error.

\subsection{The hydrogen molecule}
\label{ssec:h2}

We next define the Hamiltonian that we will consider throughout this paper. As an application of QPE, we will consider the common task of finding the ground-state energy of an electronic structure Hamiltonian. Such a Hamiltonian can be defined in second-quantized form as
\begin{equation}
    H = h_0 + \sum_{pq} h_{pq} a_p^{\dagger} a_q + \frac{1}{2} \sum_{pqrs} h_{pqrs} a_p^{\dagger} a_q^{\dagger} a_s a_r,
    \label{eq:fermionic_hamil}
\end{equation}
where $p$, $q$, $r$ and $s$ label spin orbitals. The coefficient $h_0$ defines the nuclear-nuclear contribution (which is just a number due to the Born-Oppenheimer approximation), and $h_{pq}$ and $h_{pqrs}$ are one- and two-body integrals, respectively. The form of these integrals are well known from quantum chemistry \cite{SzaboOstlund}.

In this paper we are concerned with compiling a minimal chemistry problem to lattice surgery operations, including visualization of the patch layout. We therefore consider the hydrogen molecule H$_2$ in a STO-3G basis, which is a prototypical minimal molecular example, consisting of $2$ electrons in $2$ spatial orbitals, or $4$ spin orbitals. We use an equilibrium geometry with an internuclear distance of $0.7414$ \AA.

The fermionic Hamiltonian in Eq.\ \ref{eq:fermionic_hamil} must be mapped to a qubit Hamiltonian for use in QPE. Because the minimal basis for H$_2$ consists of $4$ spin orbitals, direct mappings will result in a Hamiltonian with $4$ qubits. However, as shown by Bravyi \emph{et al}. \cite{Bravyi2017}, the qubit Hamiltonian for this problem can be reduced to just a single-qubit operator. This can be seen from symmetry arguments; the H$_2$ Hamiltonian (in this non-relativistic approximation) commutes with spin and particle-number operators, and also has spatial symmetry. Each of these  symmetries allows one qubit to be tapered. More precisely, labelling the bonding and antibonding orbitals as $\psi_g$ and $\psi_u$, and ordering the spin orbitals as $\psi_{g\uparrow}$, $\psi_{g\downarrow}$, $\psi_{u\uparrow}$, $\psi_{u\downarrow}$ (that is, using a spin-interleaved arrangement), the only determinants that contribute to the ground-state wave function are $| 1100 \rangle$ and $| 0011 \rangle$, and these two states can be represented by a single qubit. A more general approach for tapering qubits due to $\mathbb{Z}_2$ symmetries is given in Ref.~\cite{Bravyi2017}.

The Hamiltonian used takes the form
\begin{equation}
    H = c_1 Z + c_2 X,
    \label{eq:hamiltonian}
\end{equation}
with $c_1 = 0.78796736$ and $c_2 = 0.18128881$, and we have neglected the identity contribution. Note that an identical qubit Hamiltonian was considered in Ref.~\cite{Graham2022}, which performed textbook QPE on a neutral-atom quantum computer.

\subsection{Overall logical circuit}
\label{ssec:overall-logical-circuit}

Using the second-order Trotter formula techniques described in Section \ref{ssec:trotter}, we derive logical circuits for both textbook and iterative quantum phase estimation. We choose a time step $t = \pi/(c_1+c_2)$, where $c_1$ and $c_2$ are defined in Eq.\ \ref{eq:hamiltonian}, in order to ensure that eigenvalues of $H t$ are in the range $[ -\pi, \pi ]$. In this simple application, we take just a single time step in the Trotter expansion of $e^{iHt}$. We also perform QPE for just three bits of accuracy in the energy. These simplifications will of course lead to large errors in the final energy estimate; indeed, after removing rescaling factors, using three bits of precision means that the energy can only be estimated to precision $(c_1 + c_2)/4 = 0.242$ Ha. Here, we are primarily interested in understanding the required circuits in terms of lattice surgery primitives. Increasing the number of Trotter steps, or bits of precision, does not provide further insight beyond increasing the circuit depth (and number of ancilla qubits, in the case of textbook QPE).

To implement $e^{-ic_1\,Z \otimes Zt/4}$ and $e^{-ic_2\,Z \otimes Xt/2}$ we use rotation operations $R_{Z\otimes Z}$ and $R_{Z \otimes X}$, defining $R_P(\theta) = e^{-iP\theta/2}$. Thus we have rotation angles $\theta_1 = tc_1/2$ and $\theta_2 = tc_2$ for the $Z \otimes Z$ and $Z \otimes X$ rotations, respectively. We also make a minor optimisation by combining pairs of $R_{Z \otimes Z}(\theta_1)$ rotations into a single $R_{Z \otimes Z}(2\theta_1)$ rotation where possible. Figures for the logical circuits are provided in Appendix \ref{app:logical-circuit-figures}.

For both iterative and textbook QPE circuits, the gates can be grouped into three types: Pauli gates such as the $X$ gate, non-Pauli Clifford gates such as the Hadamard and $S^\dagger$ gates, and non-Clifford gates such as the two-qubit rotations and $T^\dagger$ gates. These different types of gates require different techniques to be implemented on the surface code, which we shall detail further in Section \ref{sec:implementing-logical-gates}.

\section{Implementing logical gates}
\label{sec:implementing-logical-gates}

In this section we discuss how to implement the logical circuits presented in Section \ref{ssec:overall-logical-circuit} using operations available on the surface code. The surface code represents logical qubits as patches of $d \times d$ data qubits, where $d$ is the distance of the code \cite{Kitaev2003Anyons, Dennis2002Topological}. Stabilisers consist of weight-4 $X$ and $Z$ measurements on the patch, and logical $X$ and $Z$ observables are defined along the horizontal and vertical boundaries of the patch \cite{Horsman2012}. The surface code has proven to be a popular candidate for fault tolerant quantum algorithms, due to both its high threshold and low connectivity requirements. In particular, a number of resource estimation papers use the surface code as the basis for estimating the overhead from quantum error correction \cite{Blunt2022, Kim2022LiIonBatteries, Ivanov2023SolidState, Gidney2021, litinski2023EllipticCurve}.

This section proceeds as follows. First, we approximately decompose the logical gates into a sequence of Clifford and $T$ gates. Then we consider two potential methods for implementing these gates: in Section \ref{ssec:direct-impl-clifford}, we implement the Clifford and $T$ gates directly using native lattice surgery operations; whereas in Section \ref{ssec:commute-clifford}, we use commutation relations to remove Clifford operations from the circuit, at the cost of needing to implement more general $T$-like operations.

\subsection{Decomposition to Clifford and $T$ gates}

Ideally we would want to implement logical quantum operations transversely on our error-correcting code, applying the operation to each physical qubit(s) in turn. Unfortunately, the Eastin-Knill theorem shows that this is not possible for any quantum error-correcting code \cite{Eastin2009}. In the case of the surface code, the logical gates which can be implemented transversely are single-qubit Pauli gates if the code distance is odd. Other gates within the Clifford group can be implemented on the surface code via lattice surgery operations such as patch deformation \cite{Horsman2012}, but non-Clifford gates such as the $T$ gate cannot be implemented in an error-corrected fashion.

However, it is possible to approximately decompose an arbitrary unitary operation into a sequence consisting of Clifford gates and the single-qubit $T$ gate. This was shown for arbitrary gates originally using the Solvay-Kitaev theorem \cite{Dawson2006}, and a number of improvements have been subsequently shown for both single- and multi-qubit gates \cite{Ross2016,Kliuchnikov2022}.

In the case of the QPE circuits in Section \ref{ssec:overall-logical-circuit}, the circuits contain a mixture of Clifford and non-Clifford gates. As Clifford gates can be implemented on the surface code via lattice surgery and patch-deformation techniques, such as the ones that we shall describe in Section \ref{ssec:direct-impl-clifford}, we only need to decompose the non-Clifford gates. Both the textbook and iterative QPE circuits consist of a series of two-qubit Pauli rotations as part of the Trotter expansion. In textbook QPE, there are controlled phase gates after the two-qubit rotations, to implement the inverse quantum Fourier transform. In iterative QPE, the two-qubit rotations are followed by (classically conditioned) single-qubit phase gates to implement a semi-classical version of the inverse Fourier transform.

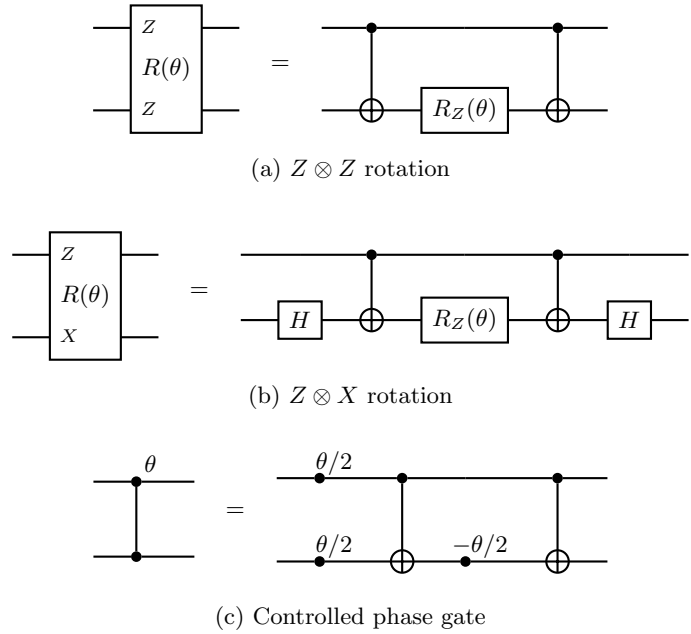
\begin{figure*}
    \begin{subfigure}[b]{\linewidth}
        \begin{quantikz}
            & \gate[2]{R(\theta)}\gateinput{$Z$} & \qw\\
            & \gateinput{$Z$} & \qw
        \end{quantikz}
        \hspace{2mm} = \hspace{1mm}
        \begin{quantikz}
            & \ctrl{1} & \qw & \ctrl{1} & \ghost[2]{R(\theta)} \qw \\
            & \targ{} & \gate{R_Z(\theta)} & \targ{} & \qw
        \end{quantikz}
        \caption{$Z \otimes Z$ rotation}
        \label{fig:zz-rotation}
    \end{subfigure}
    
    \par\bigskip
    
    \begin{subfigure}[b]{\linewidth}
        \begin{quantikz}
            & \gate[2]{R(\theta)}\gateinput{$Z$} & \qw\\
            & \gateinput{$X$} & \qw
        \end{quantikz}
        \hspace{2mm} = \hspace{1mm}
        \begin{quantikz}
            & \qw & \ctrl{1} & \qw & \ctrl{1} & \qw & \qw \\
            & \gate{H} & \targ{} & \gate{R_Z(\theta)} & \targ{} & \gate{H} & \qw
        \end{quantikz}
        \caption{$Z \otimes X$ rotation}
        \label{fig:zx-rotation}
    \end{subfigure}

    \par\bigskip
    
    \begin{subfigure}[b]{\linewidth}
        \begin{quantikz}
            & \phase{\theta} & \ghost{X} \qw\\
            & \ctrl{-1} & \ghost{X} \qw
        \end{quantikz}
        \hspace{2mm} = \hspace{1mm}
        \begin{quantikz}
            & \phase{\theta/2} & \ctrl{1} & \qw & \ctrl{1}  & \ghost{X} \qw \\
            & \phase{\theta/2} & \targ{} & \phase{-\theta/2} & \targ{} & \ghost{X} \qw
        \end{quantikz}
        \caption{Controlled phase gate}
        \label{fig:c-phase}
    \end{subfigure}
    \caption{Decompositions of parameterised (\ref{fig:zz-rotation}) $Z \otimes Z$ rotations, (\ref{fig:zx-rotation}) $Z \otimes X$ rotations, and (\ref{fig:c-phase}) controlled-phase gates into Clifford operations and single-qubit $Z$ rotations. Note that while single-qubit $Z$ rotations are equivalent to single-qubit phase gates up to a global phase, the two-qubit $Z \otimes Z$ rotation is different from a controlled phase gate due to local phases.}
    \label{fig:two-qubit-decompositions}
\end{figure*}

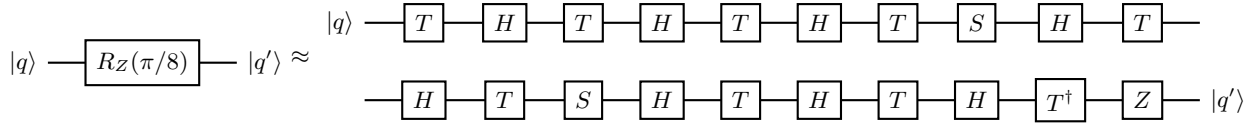
\begin{figure*}
    \begin{quantikz}
        \lstick{\ket{q}} & \gate{R_Z(\pi/8)} & \rstick{\ket{q'}} \qw
    \end{quantikz}
    $\approx$
    \begin{quantikz}
        \lstick{\ket{q}} & \gate{T} & \gate{H} & \gate{T} & \gate{H} & \gate{T} & \gate{H} & \gate{T} & \gate{S} & \gate{H} & \gate{T} & \qw\\
        & \gate{H} & \gate{T} & \gate{S} & \gate{H} & \gate{T} & \gate{H} & \gate{T} & \gate{H} & \gate{T^\dagger} & \gate{Z} & \rstick{\ket{q'}} \qw
    \end{quantikz}
    \caption{Example approximate decomposition of a $\pi/8$ $Z$ rotation into a sequence of single-qubit Clifford gates and $T$ gates using the \texttt{gridsynth} software package with three bits of precision. Global phases have been omitted, and some optimisations have been applied to combine multiple $S$ and $T$ gates.}
    \label{fig:gridsynth}
\end{figure*}

We decompose the non-Clifford gates in two steps. First, we exactly compile the two-qubit operations into Clifford gates and single-qubit $Z$ rotations and phase gates using circuit identities presented in Figure \ref{fig:two-qubit-decompositions}. Second, we use the \texttt{gridsynth} software package to approximately decompose the single-qubit $Z$ rotations into sequences of one-qubit Clifford and $T$ gates \cite{Ross2016}. Note that the two-qubit $Z \otimes Z$ rotations require a different decomposition to the controlled phase gates, due to differences in local phases. In comparison, single-qubit $Z$ rotations are equivalent to single-qubit phase gates up to a global phase $R_Z(\theta) = e^{-i\theta/2}P(\theta)$, and can therefore be decomposed using the same techniques. An example of using \texttt{gridsynth} to approximately decompose a single-qubit $Z$ rotation into the Clifford and $T$ gate set is provided in Figure \ref{fig:gridsynth}.

There is a trade-off to be made between the accuracy of decompositions generated by \texttt{gridsynth} and the number of gates required. \texttt{gridsynth} approximates a single rotation $R_Z(\theta)$ up to error $\epsilon$ in the operator norm with typically $3\log_2(1/\epsilon) + O(\log(\log(1/\epsilon)))$ non-Clifford gates \cite{Ross2016}. To get an understanding of how this extends to a whole circuit, we ran simulations of the textbook and iterative QPE circuits with \texttt{gridsynth} decompositions of varying accuracy, from 1 bit to 32 bits. For each number of bits of accuracy, we generate 1000 circuits with the single-qubit rotations decomposed to that degree of accuracy, and simulate each circuit 10,000 times.

\begin{figure*}
    \begin{subfigure}[b]{0.4\linewidth}
        \includegraphics[width=\linewidth]{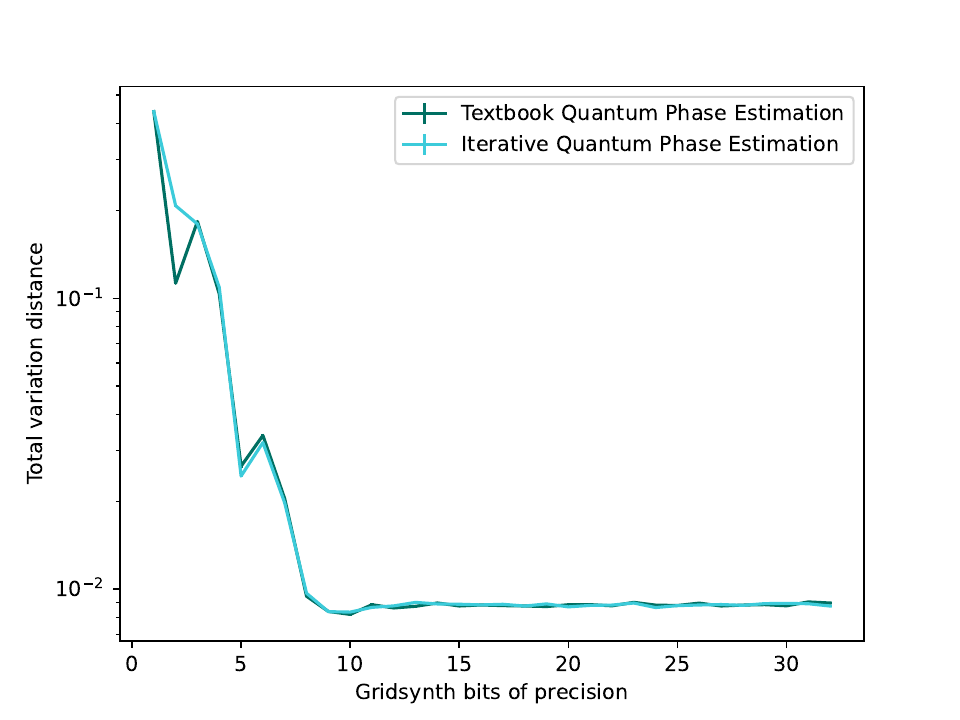}
        \caption{\label{fig:distance-per-bits-gridsynth}}
    \end{subfigure}
    \begin{subfigure}[b]{0.4\linewidth}
        \includegraphics[width=\linewidth]{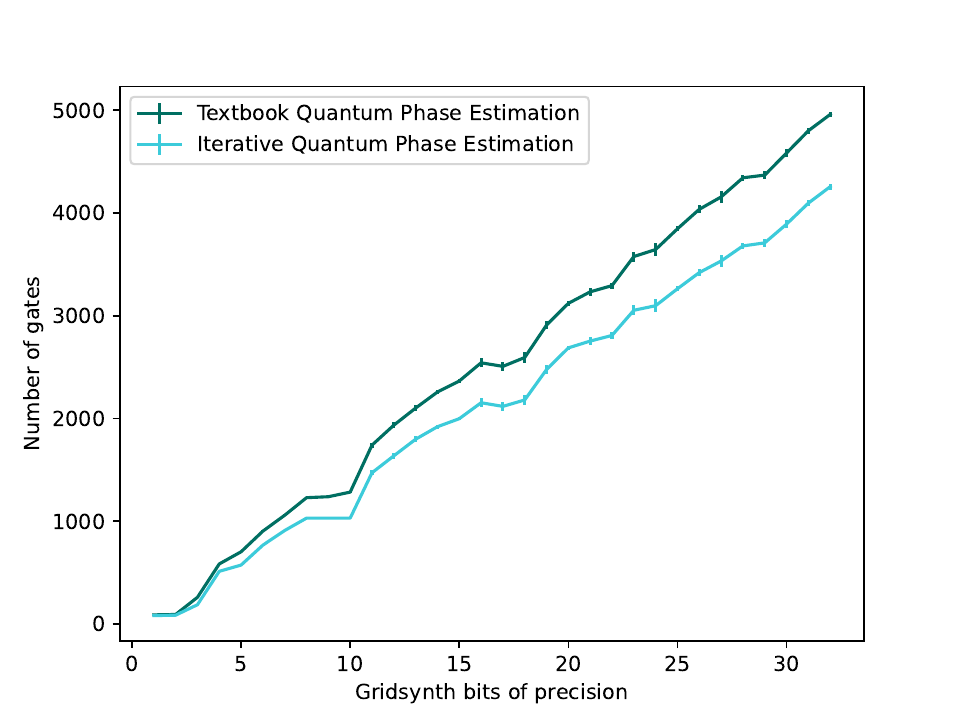}
        \caption{\label{fig:gates-per-bits-gridsynth}}
    \end{subfigure}
    \caption{Performance of \texttt{gridsynth} on textbook and iterative QPE, for increasing bits of precision in the \texttt{gridsynth} decomposition (with $3$ bits of precision used in each QPE circuit). (\ref{fig:distance-per-bits-gridsynth}) Comparison of the decomposed circuits to the exact circuits in terms of total variation distance of the output distributions. (\ref{fig:gates-per-bits-gridsynth}) The number of gates in the overall circuit.}
    \label{fig:gridsynth-trade-off}
\end{figure*}

The results are presented in Figure \ref{fig:gridsynth-trade-off}. In Figure \ref{fig:distance-per-bits-gridsynth}, we take the total variation distance between the output distributions of the decomposed circuits with that of the perfect QPE circuit. From this we see that for both textbook and iterative QPE, the total variation distance reduces quickly to approximately $8.3 \times 10^{-3}$ at 10 bits of precision per gate decomposition, but tails off beyond this value. This is due to finite precision used when estimating the total variation distance from samples. In the following results, we choose 10 bits of precision for the decomposition of phase gates, as it provides sufficient overall total variation distance for purpose of this circuit.

We also present the number of gates required for each gate decomposition accuracy in Figure \ref{fig:gates-per-bits-gridsynth}. For 10 bits of precision, there are approximately 1,300 and 1,000 logical gates for textbook and iterative QPE, respectively. Fewer logical gates can also be used at the cost of increased error; for example, fewer than 1,000 logical gates can be achieved with 5 bits of precision per rotation: 870 gates for textbook QPE, and 740 gates for iterative QPE. The total variation distance at 5 bits of precision is 2.4\%.

The results in Figure \ref{fig:gates-per-bits-gridsynth} also show that for this particular circuit iterative QPE requires fewer gates than textbook QPE regardless of decomposition accuracy. This is due to the fact that the inverse QFT step of textbook QPE requires two-qubit controlled phase rotations around fixed angles $\theta$. These are subsequently decomposed into smaller rotations $\theta/2$ and $-\theta/2$, which are then approximately decomposed using \texttt{gridsynth}. In comparison, iterative QPE works with single qubit phase rotations $\theta$ which are classically controlled. As these single angles are larger than those used for the single-qubit rotations in textbook QPE, fewer gates are required to decompose them up to a desired accuracy. For this particular QPE circuit, which is only performed to three bits of accuracy, the smallest rotation angle required beyond the Hamiltonian simulation step in the iterative QPE circuit is $-\pi/4$, which can be implemented as a single $T^\dagger$ gate. Hence for the rest of this paper we shall primarily focus on iterative QPE.

Finally, for anyone curious to see an example of the complete logical circuit, we have included example QASM circuits in the supplementary material to this paper \cite{SourceCode}, including an iterative QPE circuit with phase gates decomposed up to 10 bits of precision. This circuit features a total of 1029 operations, of which there are 13 $X$ gates, 169 $Z$ gates, 34 CNOT gates, 411 Hadamard gates, 13 $S$/$S^\dagger$ gates, 386 $T$/$T^\dagger$ gates, and three measurements in the $Z$ basis. This is the circuit we will estimate the resources for in Section \ref{sec:qec-overheads}. Note that \texttt{gridsynth} is a randomised process, and so different runs might produce different gate decompositions than presented here.

\subsection{Directly implementing Clifford and $T$ gates}
\label{ssec:direct-impl-clifford}

Next, we consider methods to implement Clifford and $T$ gates in the logical circuit. These can either be applied directly, or can be moved to the end of the logical circuit \cite{Litinski2019}. In this section we first discuss the time and space cost of directly implementing both Clifford and $T$ gates. Both of these estimates will be calculated in terms of the code distance $d$. The approach of moving Clifford operations will then be considered in Section~\ref{ssec:commute-clifford}.

The simplest gates to implement on the surface code are single-qubit Pauli gates. These operations can be implemented by either applying the corresponding Pauli gate to all data qubits if the distance $d$ is odd, or, if the distance $d$ is even, by tracking their values in software. Due to their simplicity, we shall not focus on how to implement them in this section. Likewise, preparation and measurement of a logical qubit in either the $Z$ or $X$ basis can be done in a single QEC round by preparing or measuring all data qubits in that basis. In many cases these operations can even be implemented at a cost of no additional QEC rounds, by preparing the data qubits at the start of the following round or measuring the data qubits at the end of the preceding round. Note however that preparing or measuring a logical qubit in the $Y$ basis is more complicated \cite{Gidney2023}.

It is important to note that not every Clifford gate presented in Section \ref{sec:implementing-logical-gates} will be directly implemented. Any sequence consisting of only $Z$, $S$/$S^\dagger$, and $T$/$T^\dagger$ gates can be implemented at the cost of implementing a single $T$ gate (as shown in Appendix \ref{app:implementing-t-like-gates}). Thus we can think of the sequences of gates generated by \texttt{gridsynth} such as those shown in Figure \ref{fig:gridsynth} as equivalent to sequences of alternating Hadamards and $T$-like gates.

\begin{figure}
    \includegraphics[width=0.6\linewidth]{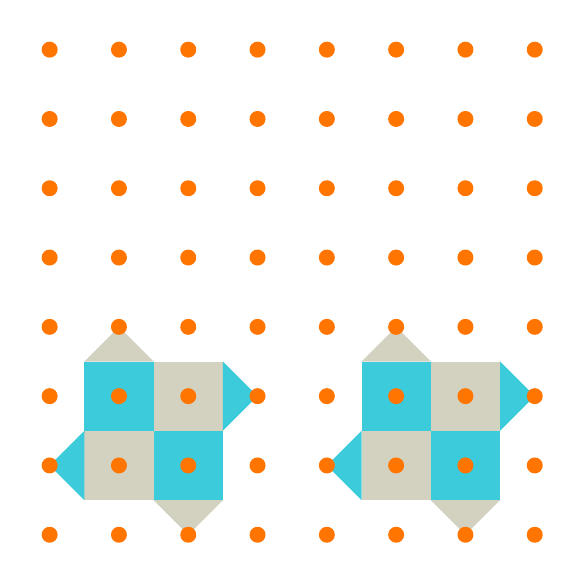}
    \caption{Layout of logical qubits as distance $d = 3$ surface code patches which can be used when directly implementing Clifford and $T$ operations. Orange dots represent qubits used for measuring stabilisers, which are represented by squares and triangles. $X$ and $Z$ stabilisers are coloured in grey and blue, respectively. Data qubits are not shown, but lie on the corners of the stabilisers. Additional qubits lie outside this space for generating states required for $T$ gates, as detailed in Section \ref{ssec:generating-magic-states}.}
    \label{fig:patch-layout-do-cliffords}
\end{figure}

Before discussing how to implement non-Pauli gates, we present how our logical qubits are arranged on a quantum processor with nearest-neighbour connectivity. For iterative QPE, we have two logical qubits, each of which is represented by a $d \times d$ patch. The primary lattice surgery operations we utilise are for implementing joint $Z \otimes Z$ measurements. We arrange our logical qubits as $d \times d$ patches such that performing joint measurements with the horizontal observable is easy. We also introduce two additional spaces of $d \times d$ data qubits, which can be used as both routing space for performing joint measurements with the vertical operator, and for additional qubits required for implementing logical gates. We have the layout in Figure \ref{fig:patch-layout-do-cliffords}, which for distance $d$ uses a total of $(2d+2)^2$ data qubits, or $2(2d+2)^2$ physical qubits including those used for measurement.

\subsubsection{CNOT gate}
\label{sssec:cnot-gate}

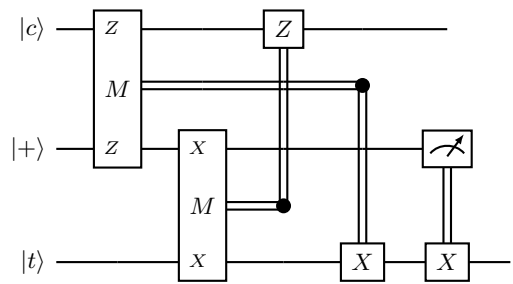
\begin{figure}
    \begin{quantikz}
        \lstick{\ket{c}} & \gate[3, nwires={2}]{M}\gateinput{$Z$} & \qw & \gate{Z} & \qw & \qw \\
        & & \cw & \cw & \cwbend{3} \\
        \lstick{\ket{+}} & \gateinput{$Z$} & \gate[3, nwires={2}]{M}\gateinput{$X$} & \qw & \qw & \meter{} \vcw{2} \\
        & & & \cwbend{-3} \\
        \lstick{\ket{t}} & \qw & \gateinput{$X$} & \qw & \gate{X} & \gate{X} & \qw
    \end{quantikz}
    \caption{Circuit for implementing a CNOT gate through joint $Z \otimes Z$ and $X \otimes X$ measurements. An additional qubit initialised in the $\ket{+}$ state is required.}
    \label{fig:cnot-circuit}
\end{figure}

A CNOT gate between a control qubit $c$ and target qubit $t$ can be implemented based on two-qubit joint Pauli measurements \cite{Horsman2012,Litinski2018}, see Figure \ref{fig:cnot-circuit}. Namely, an auxiliary qubit $a$ is initialised in the $\ket{+}$ state, followed by two joint measurements: $Z_c \otimes Z_a$ and $X_t \otimes X_a$. Finally, the auxiliary qubit is measured out in the $Z$ basis, and Pauli corrections are applied based on the outcomes.

\begin{figure*}
    \begin{subfigure}[b]{0.3\linewidth}
        \includegraphics[width=\linewidth]{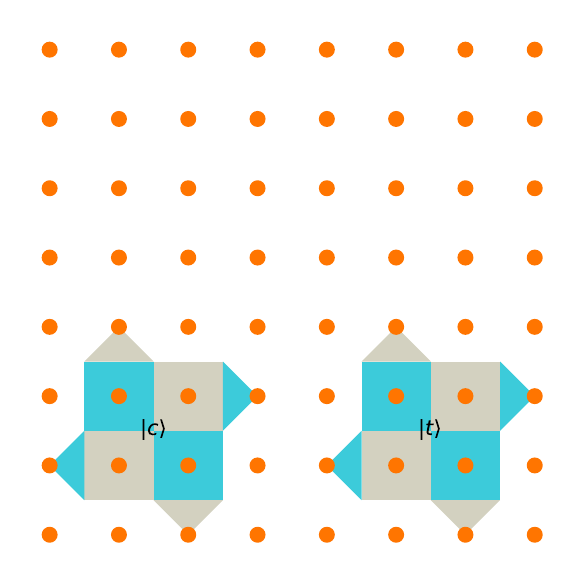}
        \caption{Initial arrangement.}
        \label{fig:cnot-start}
    \end{subfigure}
    \begin{subfigure}[b]{0.3\linewidth}
        \includegraphics[width=\linewidth]{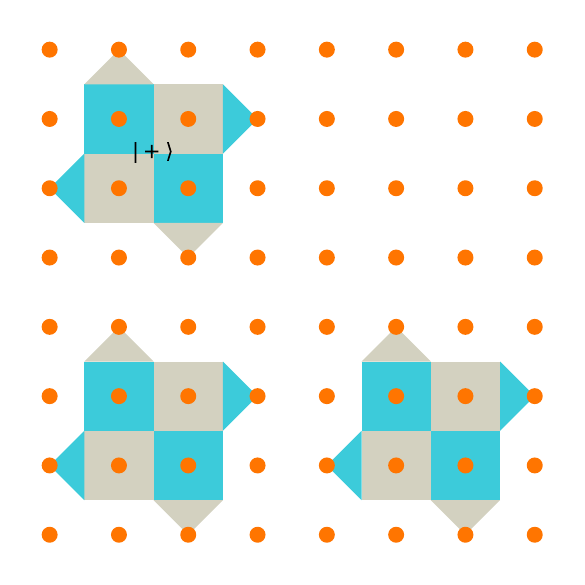}
        \caption{First stage, one QEC round.}
        \label{fig:cnot-stage-one}
    \end{subfigure}
    \begin{subfigure}[b]{0.3\linewidth}
        \includegraphics[width=\linewidth]{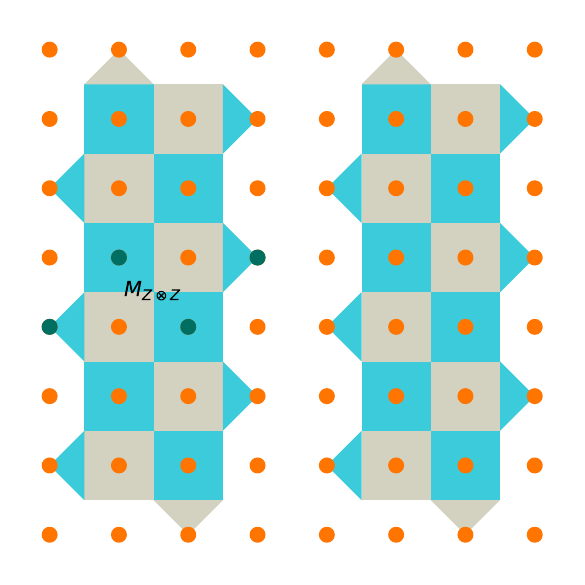}
        \caption{Second stage, $d + 1$ QEC rounds.}
        \label{fig:cnot-stage-two}
    \end{subfigure}
    \begin{subfigure}[b]{0.3\linewidth}
        \includegraphics[width=\linewidth]{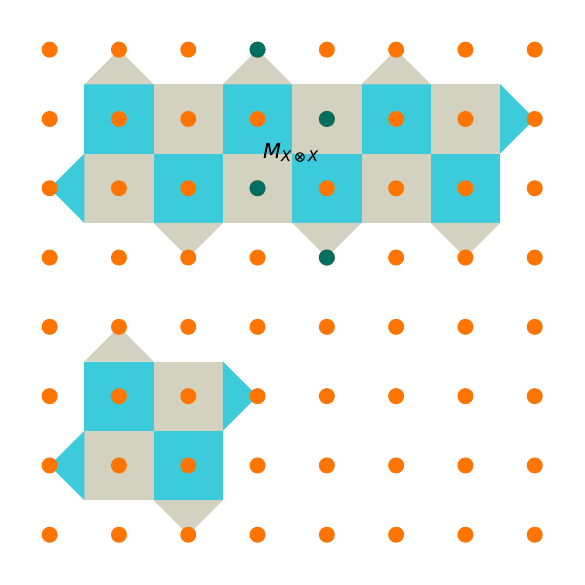}
        \caption{Third stage, $d + 1$ QEC rounds.}
        \label{fig:cnot-stage-three}
    \end{subfigure}
    \begin{subfigure}[b]{0.3\linewidth}
        \includegraphics[width=\linewidth]{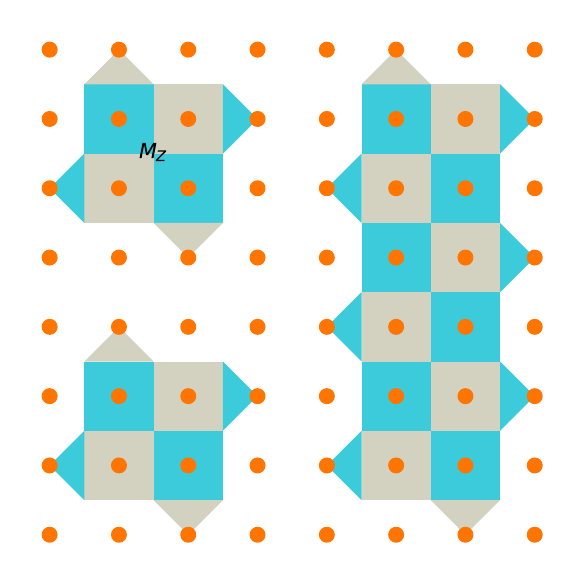}
        \caption{Final stage, $d$ QEC rounds.}
        \label{fig:cnot-stage-four}
    \end{subfigure}
    \begin{subfigure}[b]{0.3\linewidth}
        \includegraphics[width=\linewidth]{patch_do_cliffords.pdf}
        \caption{Final arrangement, one QEC round.}
        \label{fig:cnot-end}
    \end{subfigure}
        \caption{Implementation of a CNOT via lattice surgery operations, with the left patch of (\ref{fig:cnot-start}) being the control qubit and the right patch being the target qubit. Green dots represent stabiliser measurements whose outcomes produce the result of the joint logical Pauli measurement.}
    \label{fig:cnot-lattice-surgery}
\end{figure*}

This operation can be implemented on our patches via the protocol shown in Figure \ref{fig:cnot-lattice-surgery}. In Figure \ref{fig:cnot-stage-one}, we use the routing space to initialise an additional patch in the $\ket{+}$ state. We then use a merge-and-split operation between the horizontal boundaries of the control patch and the auxiliary patch to perform the $Z \otimes Z$ measurement, and at the same time grow and shrink the target patch to move it into the routing space. Next, we use another merge-and-split operation between the vertical boundaries of the target patch and the auxiliary patch to implement the $X \otimes X$ measurement. Finally, we measure out the auxiliary patch and at the same time use patch growing and shrinking to move the target qubit back to its original space. The remaining Pauli operations can either be applied transversely at the start of the next operation if the distance $d$ is odd, or simply tracked in software if the patch distance $d$ is even. The operations for growing and joining patches require $d$ QEC rounds in order to protect the code from both qubit and measurement errors, the operations for splitting and shrinking patches as well as the single-qubit logical $X$ measurement each require a single QEC round, and the Pauli operations at the end of the circuit are effectively free, meaning a total of $3d + 4$ QEC rounds are required to implement the CNOT gate.

\subsubsection{Hadamard gate}
\label{sssec:hadamard-gate}

\begin{figure*}
    \begin{subfigure}[b]{0.3\linewidth}
        \includegraphics[width=\linewidth]{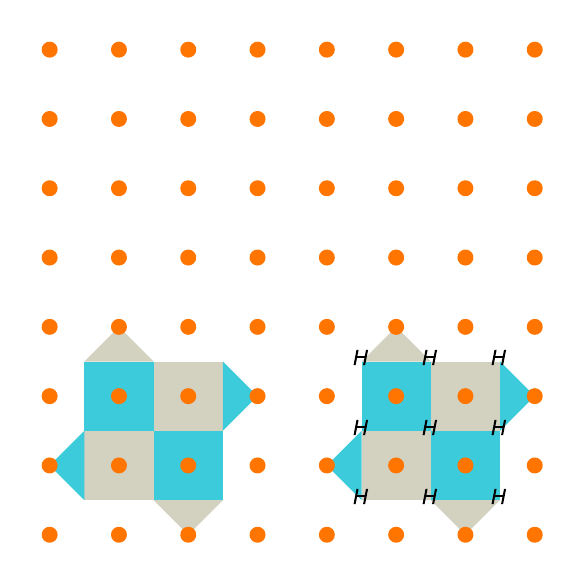}
        \caption{First stage, 0 QEC rounds.}
        \label{fig:hadamard-stage-one}
    \end{subfigure}
    \begin{subfigure}[b]{0.3\linewidth}
        \includegraphics[width=\linewidth]{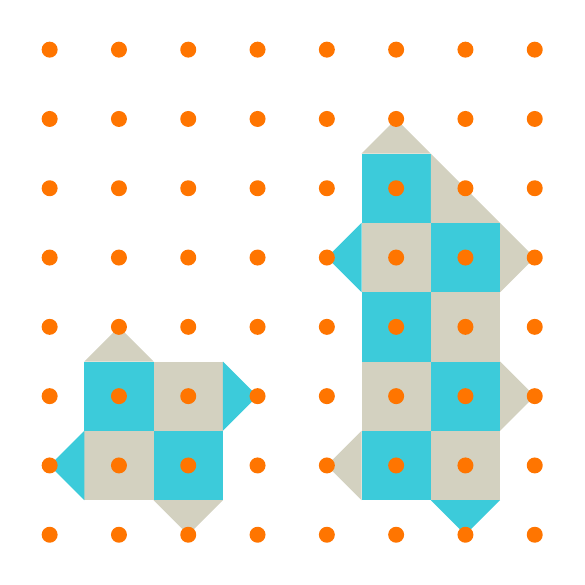}
        \caption{Second stage, $d$ QEC rounds.}
        \label{fig:hadamard-stage-two}
    \end{subfigure}
    \begin{subfigure}[b]{0.3\linewidth}
        \includegraphics[width=\linewidth]{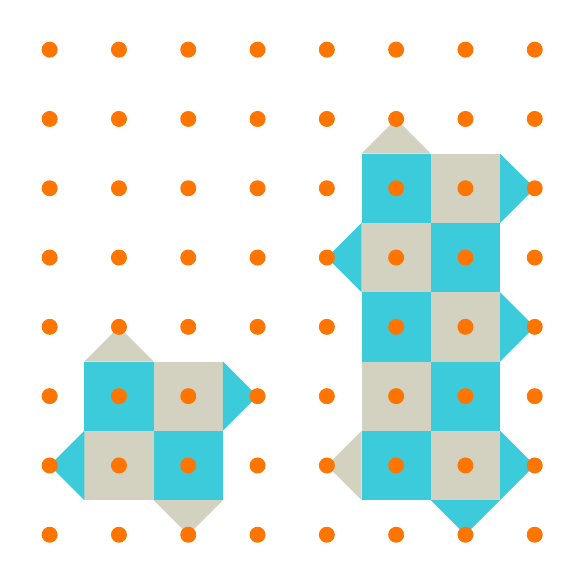}
        \caption{Third stage, $d$ QEC rounds.}
        \label{fig:hadamard-stage-three}
    \end{subfigure}
    \begin{subfigure}[b]{0.3\linewidth}
        \includegraphics[width=\linewidth]{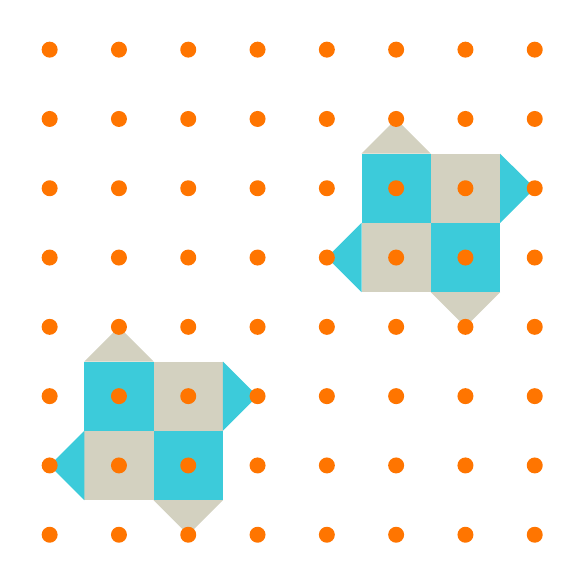}
        \caption{Fourth stage, 1 QEC round.}
        \label{fig:hadamard-stage-four}
    \end{subfigure}
    \begin{subfigure}[b]{0.3\linewidth}
        \includegraphics[width=\linewidth]{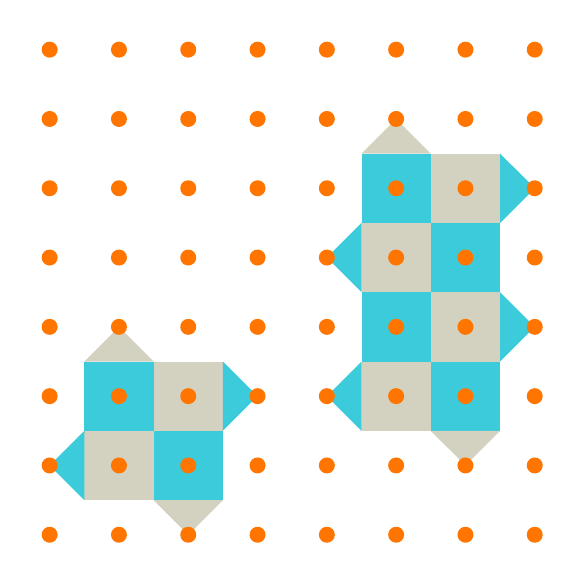}
        \caption{Fifth stage, $d$ QEC rounds.}
        \label{fig:hadamard-stage-five}
    \end{subfigure}
    \begin{subfigure}[b]{0.3\linewidth}
        \includegraphics[width=\linewidth]{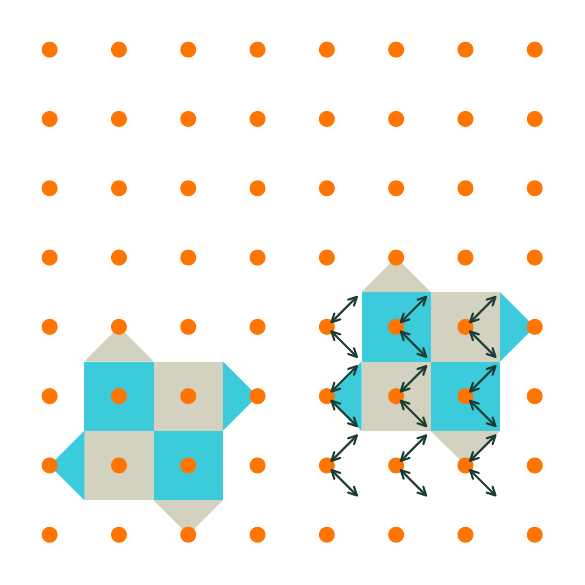}
        \caption{Final stage, 3 QEC rounds.}
        \label{fig:hadamard-stage-six}
    \end{subfigure}
        \caption{Implementation of a Hadamard gate on the right logical qubit via a transversal Hadamard gate, a series of patch deformations, and two transversal SWAP gates. Arrows denote SWAP gates between pairs of neighbouring qubits. In (\ref{fig:hadamard-stage-six}), the first QEC round is to shrink the patch, and the subsequent two QEC rounds occur after each round of SWAP gates.}
    \label{fig:hadamard-lattice-surgery}
\end{figure*}

The Hadamard gate is a Clifford gate whose role is to swap the $X$ and $Z$ observables of a qubit. Na\"{i}vely, this can be achieved on a surface code patch by applying a Hadamard operation transversely to all data qubits on the patch, as shown in Figure \ref{fig:hadamard-stage-one}. However, this has the side-effect of swapping the $X$ and $Z$ stabilisers as well as the logical observables, resulting in a different patch to the one we started with and making joint patch operations such as those used for the CNOT in Section \ref{sssec:cnot-gate} more complicated. This effect of swapping the stabilisers can be seen by comparing the patches in Figures \ref{fig:hadamard-stage-one} and \ref{fig:hadamard-stage-two}.

If we rotated the patch by 90 degrees around the central data qubit after applying the transversal Hadamard gates, then we would have implemented the logical Hadamard gate. However, this is not possible on a physical device. Instead, we use a patch deformation technique, which we present in Figures \ref{fig:hadamard-stage-two}-\ref{fig:hadamard-stage-six}, to achieve the same effect \cite{Bombin2021}. First in Figure \ref{fig:hadamard-stage-two}, we grow the patch into a longer one with length $2d + 1$. At the same time we move the corner at the top right in the original patch to the top left in the longer patch. Next in Figure \ref{fig:hadamard-stage-three}, we use patch deformation to move the corner on the bottom-right up to the top-right. At this stage the logical observables have changed directions from vertical to horizontal and vice versa. Next, we shrink the patch down in Figure \ref{fig:hadamard-stage-four}. Now we have the $X$ and $Z$ logical observables swapped, with the stabilisers in their original positions, but the whole patch has been shifted upwards.

To move this patch back to its original position, we start by growing and shrinking the patch in Figures \ref{fig:hadamard-stage-five} and \ref{fig:hadamard-stage-six}, but this leaves the patch one row of stabilisers higher than it originally was. To correct this, we use two rounds of SWAP gates to swap the data qubits with neighbouring measurement qubits, as shown in Figure \ref{fig:hadamard-stage-six}.

The most expensive parts of this process are the stages that involve patch growing and corner movement, which require $d$ QEC rounds each. Since in general two-qubit gates are much noisier than one-qubit gates, the transversal Hadamard at the start of this sequence does not require any QEC rounds. Finally, patch shrinking and transversal SWAP gates each require a single QEC round, thus requiring a total of $3d + 4$ QEC rounds.

\subsubsection{$S/S^\dagger$ gate}
\label{sssec:s-gate}

The $S$ gate, also known as the $\sqrt{Z}$ gate, is a Clifford gate that applies a phase of $i$ to the $\ket{1}$ state. Like the Hadamard gate, this gate also cannot be implemented transversely on the surface code.

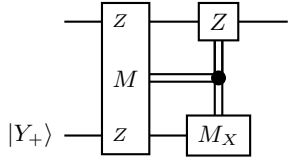
\begin{figure}
    \begin{quantikz}
        & \gate[3, nwires={2}]{M}\gateinput{$Z$} & \gate{Z} & \qw \\
        & & \cwbend{-1} & \\
        \lstick{\ket{Y_+}} & \gateinput{$Z$} & \gate{M_X} \vcw{-2} &
    \end{quantikz}
    \caption{Circuit for implementing an $S$ gate through a joint $Z \otimes Z$ measurement. An additional qubit initialised in the $\ket{Y_+} = (\ket{0} + i\ket{1})/\sqrt{2}$ state is required. Note that the $S^\dagger$ gate can be implemented by inverting the condition under which the $Z$ gate is applied.}
    \label{fig:s-circuit}
\end{figure}

There are various ways of implementing the $S$ gate using patch deformation, similarly to implementing the Hadamard in Section \ref{sssec:hadamard-gate} \cite{Bombin2021, Gidney2023}. However, these require extending the $X$ observable of a patch, and therefore require moving the patch into the routing space and back. Instead, we consider a different technique, which uses an additional patch in the $\ket{Y_+} = (\ket{0} + i\ket{1})/\sqrt{2}$ state \cite{Litinski2019}. We then perform a joint $Z \otimes Z$ measurement between this qubit and our qubit, and measure this auxiliary qubit in the $X$ basis. Finally, we apply a $Z$ correction depending on the outcomes of the two measurements. A circuit describing this operation is presented in Figure \ref{fig:s-circuit}.

\begin{figure*}
    \begin{subfigure}[b]{0.3\linewidth}
        \includegraphics[width=\linewidth]{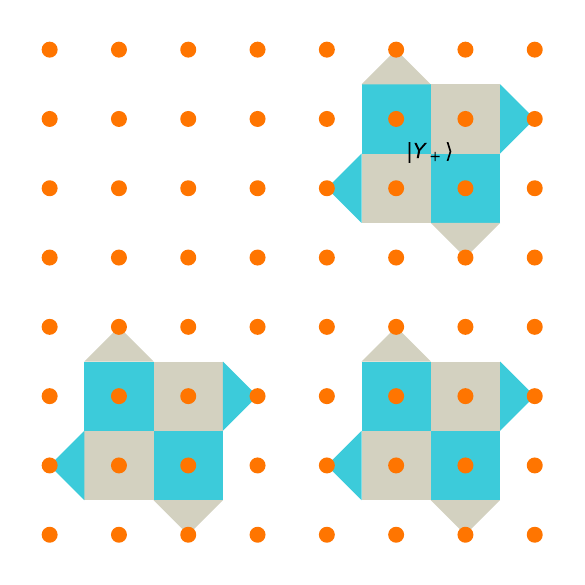}
        \caption{First stage, $d/2 + 2$ QEC rounds.}
        \label{fig:s-stage-one}
    \end{subfigure}
    \begin{subfigure}[b]{0.3\linewidth}
        \includegraphics[width=\linewidth]{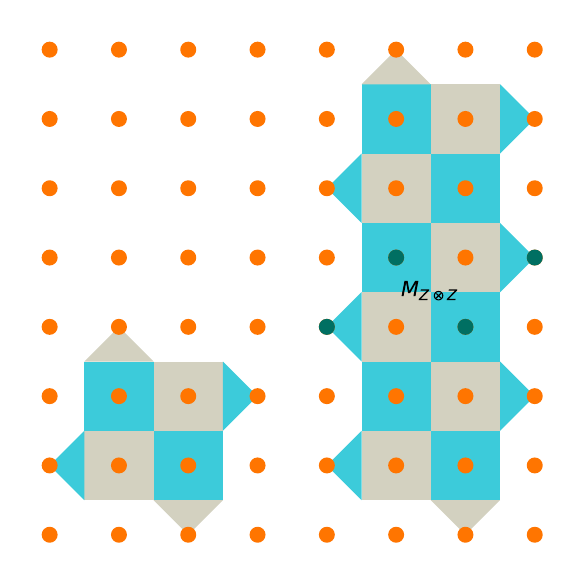}
        \caption{Second stage, $d$ QEC rounds.}
        \label{fig:s-stage-two}
    \end{subfigure}
    \begin{subfigure}[b]{0.3\linewidth}
        \includegraphics[width=\linewidth]{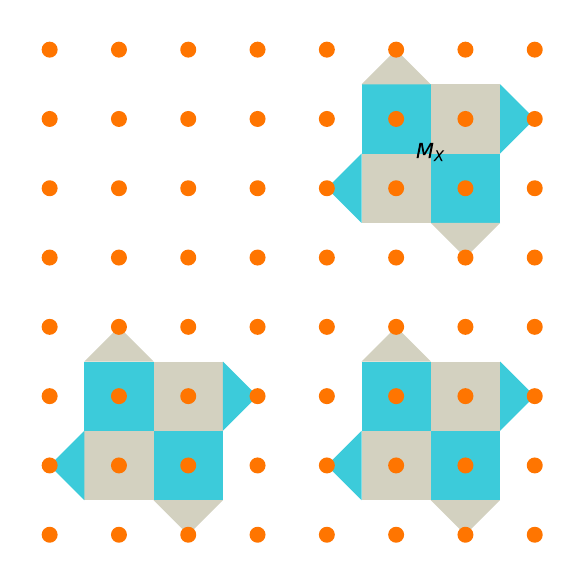}
        \caption{Third stage, one QEC round.}
        \label{fig:s-stage-three}
    \end{subfigure}
    \caption{Implementing an $S$ gate on a logical patch. In (\ref{fig:s-stage-one}), a patch in an $S$ state has been initialised in the routing space using the methods provided in Ref.~\cite{Gidney2023}. Green dots represent stabiliser measurements whose outcomes produce the result of the joint logical Pauli measurement.}
    \label{fig:s-lattice-surgery}
\end{figure*}

Note that unlike $Z$ and $X$ basis states, the $\ket{Y_+}$ state cannot be generated in a single QEC round. Instead, we utilise a different technique to generate $Y$ basis states in $d/2 + 2$ rounds with no additional qubits \cite{Gidney2023}. With this additional patch, we can implement the logical $S$ gate using the process described in Figure \ref{fig:s-lattice-surgery}. Generating the $\ket{Y_+}$ state in Figure \ref{fig:s-stage-one} takes $d/2 + 2$ QEC rounds, the joint measurement in Figure \ref{fig:s-stage-two} takes $d$ QEC rounds, and measuring the $\ket{Y_+}$ state in the $X$ basis takes a single QEC round. Any $Z$ correction can be applied in software at no additional cost, so the total number of QEC rounds required is $3d/2 + 3$. Finally, note that the $S^\dagger = SZ$ gate can also be implemented at no additional cost, by simply inverting the conditions under which the $Z$ correction is applied.

\subsubsection{$T/T^\dagger$ gate}
\label{sssec:t-gate}

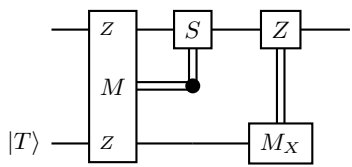
\begin{figure}
    \begin{quantikz}
        & \gate[3, nwires={2}]{M}\gateinput{$Z$} & \gate{S} & \gate{Z} & \qw \\
        & & \cwbend{-1} & & \\
        \lstick{\ket{T}} & \gateinput{$Z$} & \qw & \gate{M_X} \vcw{-2} &
    \end{quantikz}
    \caption{Circuit for implementing a $T$ gate through a joint $Z \otimes Z$ measurement. An additional qubit initialised in the $\ket{T}=(\ket{0} + e^{i\pi/4}\ket{1})/\sqrt{2}$ state is required.}
    \label{fig:t-gate-circuit}
\end{figure}

The $T$ gate is a non-trivial gate to implement on the surface code as it cannot be performed either transversely or via lattice surgery operations such as patch deformation. Instead, we introduce an auxiliary qubit initialised in the $\ket{T}=(\ket{0} + e^{i\pi/2}\ket{1})/\sqrt{2}$ state. With this $\ket{T}$ state prepared, we can implement the $T$ gate using techniques similar to those for implementing the $S$ gate in Section \ref{sssec:s-gate}. The circuit is presented in Figure \ref{fig:t-gate-circuit} \cite{Litinski2019}. First, we perform a joint $Z \otimes Z$ measurement between the data qubit and the auxiliary qubit. Next, we perform an $S$ gate conditioned on the result of this measurement outcome. Finally, we measure out the auxiliary qubit in the $X$ basis, and depending on this measurement outcome apply a final $Z$ gate to the data qubit. However, while the $\ket{Y_+}$ state can be prepared on the surface code in a fault-tolerant way in $d/2 + 2$ QEC rounds, the $\ket{T}$ state cannot be prepared on the surface code in an error-corrected fashion, and thus additional work is required in order to prepare a high-quality $\ket{T}$ state. We shall detail this further in Section \ref{ssec:generating-magic-states}.

\begin{figure*}
    \begin{subfigure}[b]{0.3\linewidth}
        \includegraphics[width=\linewidth]{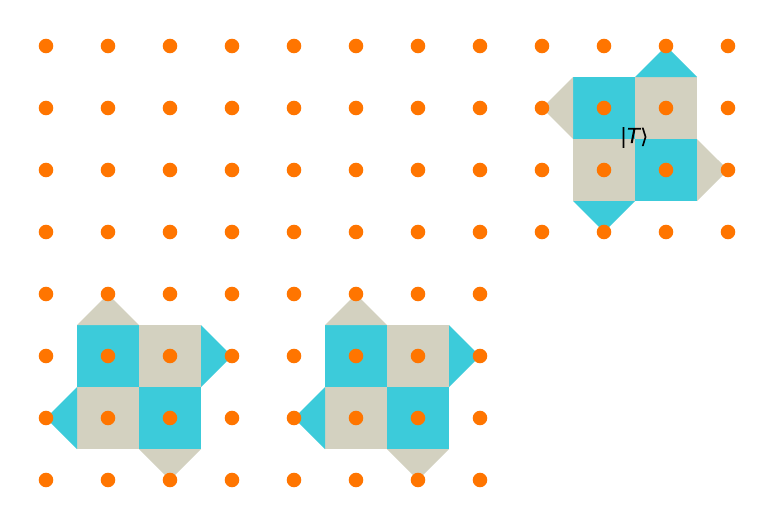}
        \caption{Initial arrangement.}
        \label{fig:t-start}
    \end{subfigure}
    \begin{subfigure}[b]{0.3\linewidth}
        \includegraphics[width=\linewidth]{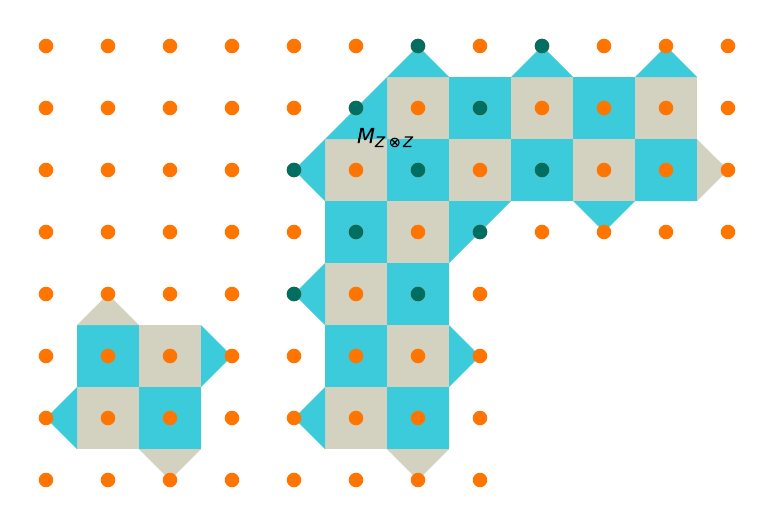}
        \caption{First stage, $d$ QEC rounds.}
        \label{fig:t-stage-one}
    \end{subfigure}
    \begin{subfigure}[b]{0.3\linewidth}
        \includegraphics[width=\linewidth]{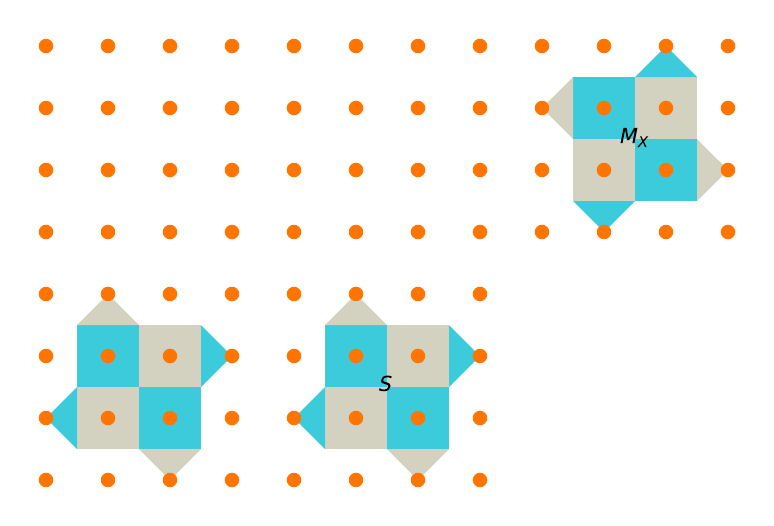}
        \caption{Second stage, $3d/2 + 4$ QEC rounds.}
        \label{fig:t-stage-two}
    \end{subfigure}
    \caption{Implementing a $T$ gate on a logical patch. In (\ref{fig:t-start}), a patch in a $T$ state has been provided in some additional space generated by a magic state factory using methods described in Section \ref{ssec:generating-magic-states}. Green dots represent stabiliser measurements whose outcomes produce the result of the joint logical Pauli measurement. $S$ correction is not shown, but occurs in (\ref{fig:t-stage-two}) after the Pauli $X$ measurement.}
    \label{fig:t-gate-lattice-surgery}
\end{figure*}

We can implement this circuit on our patch layout using the process shown in Figure \ref{fig:t-gate-lattice-surgery}. Note that the patch for the $\ket{T}$ state is not stored in the routing space like the $\ket{Y_+}$ is in Figure \ref{fig:s-lattice-surgery}. This is because unlike the $\ket{Y_+}$ state, the $\ket{T}$ state cannot be generated in a fault tolerant process, and instead needs to be generated elsewhere and stored outside of the routing space until it is required. Also note that the patch for the $\ket{T}$ state in Figure \ref{fig:t-start} is rotated compared to the patches for our data qubits, such that the vertical observable on the auxiliary patch matches the horizontal observable on our data patches. We use this to perform a joint $Z \otimes Z$ measurement between our auxiliary patch and our data patch via merge-and-split operations in Figure \ref{fig:t-stage-one}. Finally, in Figure \ref{fig:t-stage-two} we measure our auxiliary patch in the $X$ basis, and at the same time we potentially apply an $S$ correction using the methods described in Section \ref{sssec:s-gate}. As with the CNOT gate presented in Section \ref{sssec:cnot-gate}, the $Z$ operation is effectively free as it can be either tracked in software or implemented transversely. The joint measurement requires $d$ QEC rounds, the $X$ measurement requires a single QEC round, and the $S$ correction requires $3d/2 + 3$ QEC rounds, leading to a total of $5d/2 + 4$ QEC rounds to implement a logical $T$ gate.

Finally, it is worth noting that other sequences of gates can also be implemented using these techniques with no extra cost. In general, any sequence consisting of only $T$/$T^\dagger$, $S$/$S^\dagger$, and $Z$ gates can be implemented using the protocol above at the cost of implementing a single $T$ gate. Further details are provided in Appendix \ref{app:implementing-t-like-gates}.

\subsection{Moving Clifford gates}
\label{ssec:commute-clifford}

In this section we will consider another way of implementing the logical circuit from Section \ref{sec:logical-circuit} on the surface code based on \cite{Litinski2019}. This technique offers the benefit of only needing to think about how to implement the non-Clifford gates, but at the cost of increasing the complexity of implementing such gates.

\subsubsection{Pauli product rotations}

The key to this implementation method is that the logical gates we want to implement can be realised as rotations in a particular single- or multi-qubit Pauli basis. More formally, an $n$-qubit quantum gate can be implemented as a sequence of rotations $R_{P_j}(\theta_j) = e^{-iP_j\theta_j/2}$ for suitably chosen $P_j \in \{I, X, Y, Z\}^{\otimes n}$ and $\theta_j$ \footnote{An astute reader may notice that our definition of $R_{P_j}(\theta_j)$, and by extension our translation of Clifford and non-Clifford gates to Pauli rotations, differs from that in Ref.~\cite{Litinski2019} by a factor of 2. This is to ensure correct periodicity, such that $R_P(\theta + 2\pi) = R_P(\theta) \forall P, \theta$. This is also consistent with the definition of Pauli rotations in other texts such as, for example, \cite{Nielsen2010}}. The simplest example of this phenomenon is the Pauli-gates themselves, which can be implemented as $P = R_P(\pi)$. Similarly, the $T$ and $S$ gates are both single-qubit rotations in the $Z$ basis, and can thus be realised as $T = R_Z(\pi/4)$ and $S = R_Z(\pi/2)$, respectively. Single-qubit Pauli measurements, although not rotations around a Pauli basis, can also be seen as operations which project a state into a Pauli basis. In the case of QPE for example, measurements project a state into the $Z$ basis.

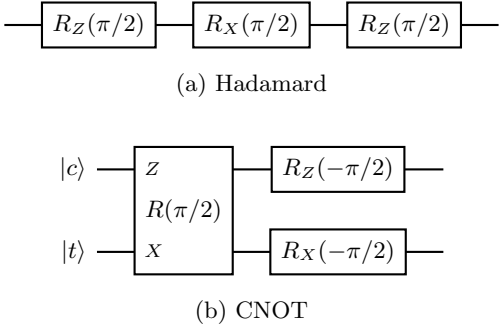
\begin{figure}
    \begin{subfigure}[b]{\linewidth}
        \begin{quantikz}
            & \gate{R_Z(\pi/2)} & \gate{R_X(\pi/2)} & \gate{R_Z(\pi/2)} & \qw
        \end{quantikz}
        \caption{Hadamard}
        \label{fig:hadamard-pauli-rotation}
    \end{subfigure}
    
    \par\bigskip
    
    \begin{subfigure}[b]{\linewidth}
        \begin{quantikz}
            \lstick{\ket{c}} & \gate[2]{R(\pi/2)}\gateinput{$Z$} & \gate{R_Z(-\pi/2)} & \qw \\
            \lstick{\ket{t}} & \gateinput{$X$} & \gate{R_X(-\pi/2)} & \qw
        \end{quantikz}
        \caption{CNOT}
        \label{fig:cnot-pauli-rotation}
    \end{subfigure}
    \caption{Implementing (\ref{fig:hadamard-pauli-rotation}) Hadamard and (\ref{fig:cnot-pauli-rotation}) CNOT gates as Pauli $\pi/2$ rotations.}
    \label{fig:clifford-pauli-rotation}
\end{figure}

The remaining gates to translate into this picture are the CNOT and Hadamard gates. Although not as easy to see as the gates listed above, both of these gates can be implemented as sequences of Pauli $\pi/2$ rotations given in Figure \ref{fig:clifford-pauli-rotation} \cite{Litinski2019}. The Hadamard gate can be decomposed as $H = R_Z(\pi/2) \cdot R_X(\pi/2) \cdot R_Z(\pi/2)$, up to a global phase. The CNOT can be written as a joint $\pi/2$ $Z \otimes X$ rotation, followed by a $-\pi/2$ $Z$ rotation on the control qubit, and a $-\pi/2$ $X$ rotation on the target qubit. This is similar to the circuit used in Figure \ref{fig:cnot-circuit}, but with Pauli $\pi/2$ rotations rather than Pauli measurements.

\subsubsection{Moving Pauli rotations}
\label{sssec:commute-pauli-rotations}

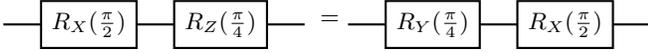
\begin{figure}
    \hspace{-0.6cm}
    \begin{quantikz}
        & \gate{R_X(\frac{\pi}{2})} & \gate{R_Z(\frac{\pi}{4})} & \qw
    \end{quantikz}
    \hspace{0.1cm}=\hspace{-0.1cm}
    \begin{quantikz}
        & \gate{R_Y(\frac{\pi}{4})} & \gate{R_X(\frac{\pi}{2})} & \qw
    \end{quantikz}
    \caption{Moving a $\pi/2$ $X$ rotation past a $\pi/4$ $Z$ rotation.}
    \label{fig:commuting-x-z}
\end{figure}

The benefit of describing operations as rotations in a Pauli basis is that it becomes easier to understand how to transform them without modifying the outcome of the circuit. For example, in Figure \ref{fig:commuting-x-z}, a $\pi/2$ rotation in the $X$ basis is moved past a $\pi/4$ rotation in the $Z$ basis. The result is that the $Z$ rotation is transformed into a $\pi/4$ rotation in the $iXZ = i(-iY) = Y$ basis.

These transformations can be applied more generally as well, the rules for which we discuss in Appendix \ref{app:commutation-rules}. The benefit of these transformations to the circuit is that we can move all $\pi$ and $\pi/2$ Pauli rotations, which correspond to Pauli and Clifford operations, past the final measurement operation of the circuit. Operations beyond this point do not affect the outcome of our circuit, and therefore do not need to be implemented. Thus we have reduced our circuit to only involving $\pi/4$ Pauli rotations, which correspond to a generalisation of $T$ gates, and joint Pauli measurements. We shall now look at how to implement these more general operations.

\subsubsection{Implementing $\pi/4$ joint Pauli rotations}

First we shall show how to reduce the $\pi/4$ joint Pauli rotations to joint Pauli measurements. These will then be implemented using a particular patch layout and lattice surgery operations in Section \ref{sssec:joint-pauli-measurements}.

\begin{figure}
    \begin{quantikz}
        \lstick{\ket{\Psi}} & \gate[3, nwires={2}]{M}\gateinput{$P$} \qwbundle[alternate]{} & \gate{R_P(\pi/2)} \qwbundle[alternate]{} & \gate{R_P(\pi)} \qwbundle[alternate]{} & \qwbundle[alternate]{} \\
        & & \cwbend{-1} & & \\
        \lstick{\ket{T}} & \gateinput{$Z$} & \qw & \gate{M_X} \vcw{-2} &
    \end{quantikz}
    \caption{Implementing a $\pi/4$ rotation on an $n$-qubit quantum state $\ket{\Psi}$ in a Pauli basis $P$ using an auxiliary state $\ket{T} = (\ket{0} + e^{i\pi/4}\ket{1})/\sqrt{2}$. This can be seen as a generalisation of the $T$ gate circuit in Figure \ref{fig:t-gate-circuit}.}
    \label{fig:pi-4-circuit}
\end{figure}
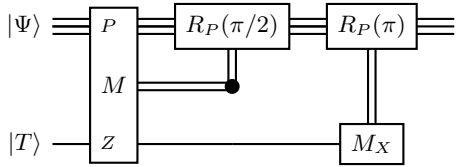

A circuit for implementing $\pi/4$ rotations is presented in Figure \ref{fig:pi-4-circuit}. This can be seen as a generalisation of the $T$ gate circuit in Figure \ref{fig:t-gate-circuit}, where now the single-qubit $Z$ basis has been replaced with a general multi-qubit basis $P$. The auxiliary qubit required for this operation is the same $\ket{T} = (\ket{0} + e^{i\pi/4}\ket{1})/\sqrt{2}$ state from Section \ref{sssec:t-gate}.

Because the rotation basis has generalised, so too have the corrective gates performed after the measurement. Now, instead of single-qubit $S$ and $Z$ gates we have more general $\pi/2$ and $\pi$ rotations in an arbitrary Pauli basis $P$. The implementation of the $\pi$ rotation is still a Pauli operation, and can be either tracked in software or implemented transversely as before. As for the $\pi/2$ rotation, one can account for this by employing the same techniques as described in Section \ref{sssec:commute-pauli-rotations} in an online fashion, moving the rotation past the final round of measurements to effectively remove it from the circuit and adjusting the subsequent operations accordingly \cite{Litinski2019}.

\subsubsection{Implementing joint Pauli measurements}
\label{sssec:joint-pauli-measurements}

\begin{figure*}
    \begin{subfigure}[b]{0.3\linewidth}
        \includegraphics[width=\linewidth]{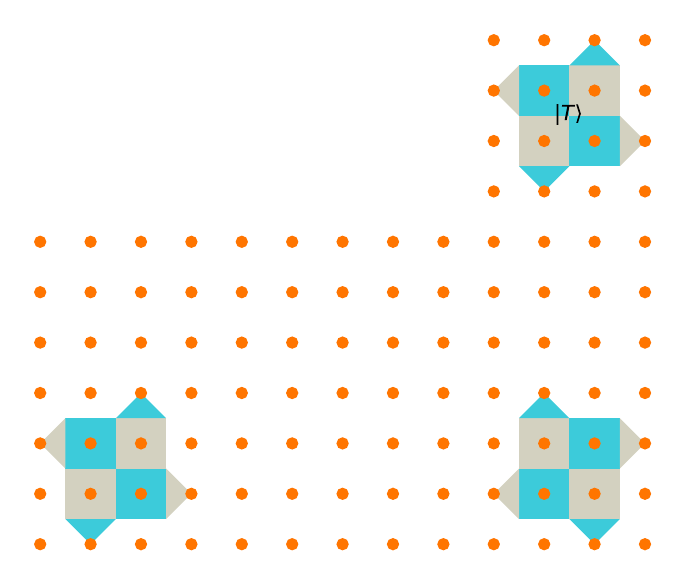}
        \caption{Initial layout.}
        \label{fig:patch-pauli-rotation-start}
    \end{subfigure}
    \begin{subfigure}[b]{0.3\linewidth}
        \includegraphics[width=\linewidth]{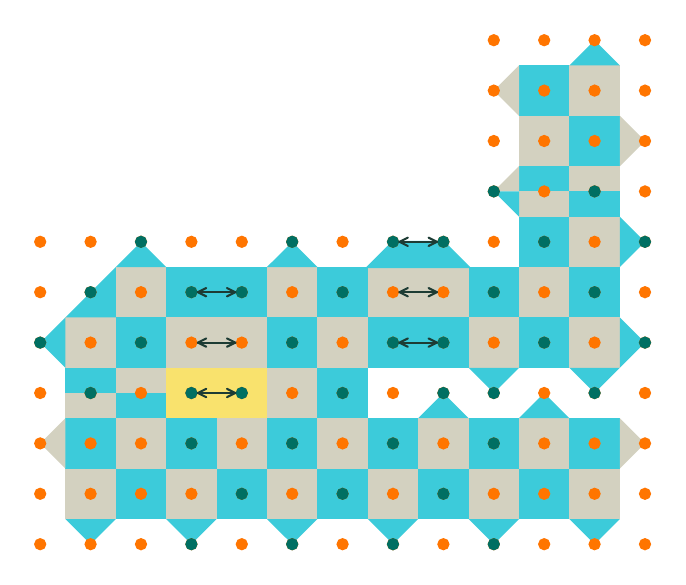}
        \caption{Pauli $Y \otimes X \otimes Z$ measurement.}
        \label{fig:patch-pauli-rotation-yx}
    \end{subfigure}
    \begin{subfigure}[b]{0.3\linewidth}
        \includegraphics[width=\linewidth]{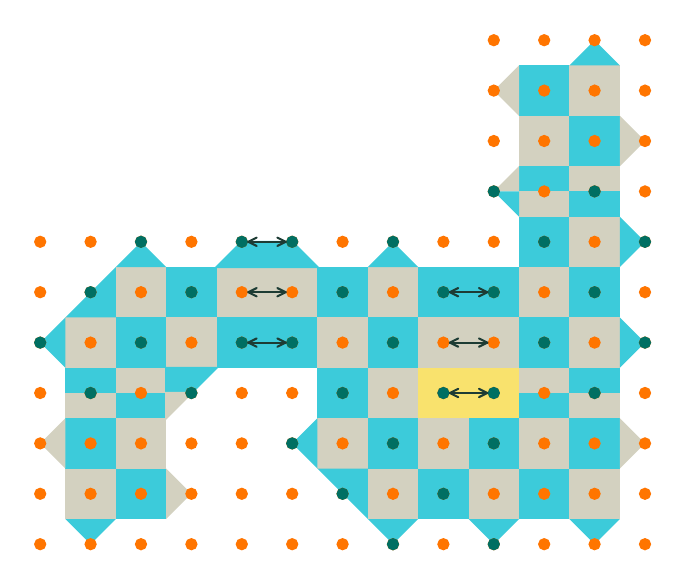}
        \caption{Pauli $Z \otimes Y \otimes Z$ measurement.}
        \label{fig:patch-pauli-rotation-zy}
    \end{subfigure}
    \caption{Layout of logical patches for implementing joint Pauli measurements. In (\ref{fig:patch-pauli-rotation-start}), the two qubits used in the logical circuit are at the bottom, and an auxiliary qubit is initialised in the $\ket{T}$ state at the top. Example joint measurements required for implementing $\pi/4$ Pauli $Y \otimes X$ and $Z \otimes Y$ operations in $d+1$ QEC rounds are presented in (\ref{fig:patch-pauli-rotation-yx}) and (\ref{fig:patch-pauli-rotation-zy}), respectively. The auxiliary $\ket{T}$ state is always measured in the $Z$ basis as part of the joint measurement. Green dots represent stabiliser measurements whose outcomes produce the result of the joint logical Pauli measurement. Twist defects are presented in yellow. Arrows between neighbouring measurement qubits show extended connectivity than what is required for the methods presented in Section \ref{ssec:direct-impl-clifford}.}
    \label{fig:multi-qubit-measurements}
\end{figure*}

Finally we discuss how to implement general Pauli measurements between patches on a surface code. The specific arrangement we use is given in Figure \ref{fig:patch-pauli-rotation-start}. Note that this patch has more routing space than the one in \ref{fig:patch-layout-do-cliffords}, this is because the more general operations require access to both the horizontal and vertical observables of the patches. This results in six logical patches arranged on a grid of $(3d + 4) \times (2d + 2)$ data qubits, or $2(3d + 4) \times (2d + 2)$ physical qubits total.

The most challenging operations to implement are those which include the $Y$ basis of a qubit. This is because the $Y$ basis does not correspond to the horizontal or vertical observable on a surface code patch, but is instead a product of both the horizontal and vertical observables. One option is to decompose $\pi/4$ rotations which involve the $Y$ basis of a qubit into a sequence of $\pi/4$ and $\pi/2$ rotations which only act on the $X$ and $Z$ bases \cite{Litinski2019}. However, doing so introduces $\pi/2$ rotations which cannot be moved past the $\pi/4$ rotation without reintroducing the $Y$ basis, so such rotations would need to be implemented.

Instead, we utilise another technique from \cite{Chamberland2022} to implement $Y$ basis measurements directly via lattice surgery operations. Some example measurements for implementing Pauli $\pi/4$ rotations in the $Y \otimes X$ and $Z \otimes Y$ bases are given in Figure \ref{fig:multi-qubit-measurements}. These joint measurements require $d$ QEC rounds, followed by a single QEC round to measure the auxiliary patch in the $X$ basis. These two sets of measurement results give us the corrections to move past future operations.

Here we utilise some lattice surgery techniques not used in Section \ref{ssec:direct-impl-clifford}. First, we add weight-five stabilisers, known as twist defects, which involve a Y Pauli term on one of the qubits. To ensure the surrounding stabilisers commute with the twist defects, we utilise two other lattice surgery techniques: first, we add domain walls, which are denoted by half-blue-half-grey squares and act as a combination of $X$ and $Z$ stabilisers; and second, we add elongated weight-four stabilisers, which are denoted by blue and grey rectangles. It is important to note that although these techniques allow for direct implementation of joint measurements involving the $Y$ basis, there is an additional cost in that measuring these longer stabilisers requires additional connectivity compared to the layout used in Section \ref{ssec:direct-impl-clifford}. These extra connections between measurement qubits are not uniform, and shown by arrows in Figure \ref{fig:multi-qubit-measurements}. In general, for distance $d$ a total of $4d$ extra connections are required for implementing this algorithm, which connect four columns of adjacent measurement qubits.

\section{Error correction overheads}
\label{sec:qec-overheads}

We are now ready to discuss the cost of implementing these logical gates on the surface code. There are two primary sources of error which contribute to the probability of a failure at the error-correction level: first, errors from generating $\ket{T}$ states, which we shall explore in Section \ref{ssec:generating-magic-states}; and second, errors from a logical failure on a qubit, which we shall explore in Section \ref{ssec:estimating-code-distance}.

\subsection{Generating $\ket{T}$ states}
\label{ssec:generating-magic-states}

Both of the methods used in Section \ref{sec:implementing-logical-gates} require additional qubits initialised in the $\ket{T}$ state. It is possible to initialise a surface code patch into an arbitrary qubit state $\ket{\psi}$, by initialising one data qubit of the patch in the $\ket{\psi}$ state, followed by $d$ rounds of measurements \cite{Horsman2012}. However, initialising a data qubit into an arbitrary state means that this qubit is initially unprotected from errors, so this method cannot be implemented in a way that reduces the logical error probability below the physical error probability. In fact, it can be shown that there is no fault-tolerant way of initialising non-stabiliser states such as the $\ket{T}$ state on the surface code \footnote{Note that there are other ways of initialising a $\ket{T}$ state on the surface code which are more immune to errors \cite{Li2015PostselectedMagicStates}, however these rely on post-selection and therefore might introduce additional overheads. For simplicity we shall not focus on this method.}.

Even though patches cannot be initialised in the $\ket{T}$ state in a way that suppresses errors, it is possible to use distillation protocols to reduce the error probability of $\ket{T}$ states. These protocols take multiple noisy $\ket{T}$ states and output a smaller number of $\ket{T}$ states with a reduced error probability \cite{Bravyi2005, Bravyi2012, Jones2013, Fowler2013, Haah2018, Litinski2019, Litinski2019magicstate, Gidney2019}. For example, if it is possible to generate 15 $\ket{T}$ states each with error probability $p$, it is possible to distill these into a single $\ket{T}$ state with error probability $35p^3$ \cite{Bravyi2005}. It is also possible to concatenate these factories to reduce the error probability even further. For example, if the $15$-to-$1$ protocol is used to generate 15 $\ket{T}$ states each with error probability $35p^3$, these can then be used in another $15$-to-$1$ protocol to generate a single $\ket{T}$ state with error probability $35(35p^3)^3 = 1,500,625p^9$ \cite{Litinski2019}. The cost with these protocols is that reducing the error probability requires additional resources in terms of both time and number of qubits. A summary of several protocols and their associated costs is provided in Ref.~\cite{Litinski2019magicstate}. We also provide some example resource estimates for $15$-to-$1$ factories in Table \ref{tab:t-factories}, generated using code from Ref.~\cite{Litinski2019magicstate}.

When choosing a suitable protocol, there are multiple factors that we need to consider. First, we need to consider the overall logical failure probability from faulty $\ket{T}$ state generation. This means that if our logical circuit uses $m$ $T$ gates -- and therefore requires $m$ $\ket{T}$ states -- we need to choose a probability of distilled state failure $p_{\text{dist}}$ such that $m \times p_{\text{dist}}$ is within our error bounds.

\begin{table}
    \caption{\label{tab:t-factories} Resource estimates for some example $15$-to-$1$ $\ket{T}$ state factories at physical error rates $10^{-3}$ and $10^{-4}$.}
    
    \begin{ruledtabular}
        \begin{tabular}{ccc}
             Physical error probability & $10^{-3}$ & $10^{-4}$ \\\hline
             $X$ error distance & $11$ & $5$ \\
             $Z$ error distance & $5$ & $3$ \\
             Measurement error distance & $5$ & $3$ \\
             Distillation error & $8.66 \times 10^{-6}$ & $4.68 \times 10^{-6}$ \\
             Number of qubits & $2066$ & $522$ \\
             Expected number of QEC rounds & $31.30$ & $18.05$ \\
        \end{tabular}
    \end{ruledtabular}
\end{table}

The second aspect we need to consider is the time required to generate each $\ket{T}$ state. In order to avoid logical qubits remaining idle as we wait for $\ket{T}$ states to be generated, we need to ensure that $\ket{T}$ states are generated fast enough that they are available as and when they are needed. This depends on both the number of QEC rounds required to generate the $\ket{T}$ states, but also the number of QEC rounds required to implement these logical operations. If we implement Clifford and $T$ gates directly as described in Section \ref{ssec:direct-impl-clifford}, the circuit primarily consists of alternating sequences of Hadamard gates, which take $3d + 4$ QEC rounds, and $T$-like gates, which take between $d + 1$ and $5d/2 + 4$ QEC rounds, depending on whether or not an $S$ gate correction is required. This means that when implementing Clifford and $T$ gates directly, a $\ket{T}$ state needs to be produced at least once every $4d + 5$ QEC rounds. In comparison, when Clifford operations have been moved through the circuit as described in Section \ref{ssec:commute-clifford}, the only operations required are a single joint Pauli measurement and a single $X$ basis measurement, meaning that a $\ket{T}$ state must be produced every $d+1$ QEC rounds. If a single distillation protocol cannot generate states fast enough, multiple instances of the protocol can be run in parallel to generate states more frequently, at the cost of increasing the number of physical qubits \cite{Litinski2019}. As we show in Appendix \ref{app:factory-arrangements}, up to four factories can be placed around the two corners at the top of the routing space. It is possible to add even more factories beyond these four, but doing so could require additional space for routing and storage of $\ket{T}$ states. On the other hand, if a logical $\ket{T}$ state can be generated faster than required, additional storage space is required to protect the state from errors while it waits to be consumed, which can be included as part of the routing space estimates.

\subsection{Estimating code distance}
\label{ssec:estimating-code-distance}

To reduce the probability of a logical error occurring on one of our logical qubits, we can tweak the code distance $d$. A higher distance will reduce the probability of getting a sequence of physical errors which lead to a logical error, but comes at the cost of increasing both the number of physical qubits per logical qubit, and the number of QEC rounds per logical operation. In the case of the surface code, the probability of a logical error on a single logical qubit per code cycle assuming a depolarising noise model can be estimated as
\begin{equation}
    p_L(p, d) = 0.1(100p)^{(d+1)/2},
    \label{eq:logical-error-rate}
\end{equation}
where $p$ is the physical error probability \cite{Fowler2013, Fowler2018, Litinski2019}. For the purpose of this application, we want to choose a sufficiently high $d$ that the probability of a logical error occurring on any qubit during any QEC round is within our error bound. We use Eq.\ \ref{eq:logical-error-rate} to approximate our probability of a logical error at any point in the computation as
\begin{equation}
    (n_{\text{data}} + n_{\text{route}}) \times n_{\text{meas}} \times p_L(p, d),
    \label{eq:distance-bound}
\end{equation}
where $n_{\text{data}}$ is the number of surface code patches for our data qubits, $n_{\text{route}}$ is the number of additional patches used for routing \footnote{Note that at various points of the computation these routing patches are unused. This means that errors occurring on them will not lead to an overall failure. However, in the worst case a logical error occurs on one of these patches while it is in use, which can in turn lead to a failure of the overall computation, hence why we include both data patches and routing patches in this calculation.}, and $n_{\text{meas}}$ is the number of QEC rounds. Given these parameters and physical error probability $p$, we can pick a distance by choosing an appropriate $d$ such that Eq.\ \ref{eq:distance-bound} is within our target failure probability.

\subsection{Results}
\label{ssec:results}

We are now ready to estimate error correction overheads for our iterative quantum phase estimation circuit. As a recap, our circuit consists of 13 $X$ gates, 169 $Z$ gates, 34 CNOT gates, 411 Hadamard gates, 13 $S$/$S^\dagger$ gates, 386 $T$/$T^\dagger$ gates, and three $Z$ basis measurements. As previously described, $X$ and $Z$ gates are free as they can be implemented transversely at the start of a QEC round. Of the $S$ and $S^\dagger$ gates, one is used in a sequence of $T$ gates, and can therefore be implemented as a $T$-like gate. This leaves our costing as 411 Hadamard gates, 34 CNOT gates, 386 $T$-like gates, 12 $S$/$S^\dagger$ gates, and three measurements.

We also need to make assumptions on the error correction requirements of our algorithm. We assume physical errors correspond to depolarising noise with a physical error probability ranging between $10^{-4}$ and $2\times 10^{-3}$. We also assume a target failure probability of $1\%$, though a higher target probability can be used to reduce overheads \cite{Gidney2019, Gidney2021}. This target failure probability is split evenly, so the probability of errors occurring from faulty $\ket{T}$ state preparation is at most $0.5\%$, and the probability of logical errors happening on the qubits used in the logical circuit is also at most $0.5\%$. For 386 $T$-like gates, the required error rate per $T$ gate in order to meet this error budget is $1.3 \times 10^{-5}$. This is a higher error probability than what is seen from many distillation techniques \cite{Litinski2019magicstate, Gidney2019}, so instead we use code from \cite{Litinski2019magicstate} to look for smaller factories which still fit within our target failure probability. Note that both factories presented in Table \ref{tab:t-factories} suffice at error rates $10^{-3}$ and $10^{-4}$.

Our results are presented in Fig.\ \ref{fig:results-summary}. To help explain these resource estimates, the rest of this section will provide detailed costings for physical error rates of $10^{-3}$ and $10^{-4}$. These physical error rates are commonly used when estimating the resource requirements of fault-tolerant quantum algorithms \cite{Litinski2019, Litinski2019magicstate, Blunt2022}. For ease of reading, a summary of these results is presented in Table \ref{tab:detailed-results}.

\begin{table*}
    \caption{\label{tab:detailed-results} Detailed resource estimates required to perform the iterative QPE circuit described in the main text for the hydrogen molecule, considering physical error rates of $10^{-3}$ and $10^{-4}$. Resource estimates for the $\ket{T}$ state factories are in Table \ref{tab:t-factories}.}
    \begin{ruledtabular}
        \begin{tabular}{ccccc}
             Implementation method & \multicolumn{2}{c}{Direct implementation} & \multicolumn{2}{c}{Move Cliffords} \\
             Physical error rate & $10^{-3}$ & $10^{-4}$ & $10^{-3}$ & $10^{-4}$ \\\hline
             Code distance & 12 & 6 & 11 & 5 \\
             Number of qubits for logical circuit & 1,352 & 392 & 1,776 & 456 \\
             Probability of logical error on any logical qubit & $4.8 \times 10^{-3}$ & $8.6 \times 10^{-4}$ & $4.2 \times 10^{-3}$ & $2.3 \times 10^{-3}$ \\
             Number of QEC rounds between non-Clifford gates & 53 & 29 & 12 & 6 \\
             Number of $\ket{T}$ state factories & 1 & 1 & 3 & 4 \\
             Number of qubits for generating $\ket{T}$ states & 2,066 & 522 & 6,198 & 2,088 \\
             Number of qubits for storing $\ket{T}$ states & 288 & 72 & 726 & 200 \\
             Probability of $\ket{T}$ state error & $3.1 \times 10^{-3}$ & $1.8 \times 10^{-3}$ & $3.1 \times 10^{-3}$ & $1.8 \times 10^{-3}$ \\\hline
             Total number of physical qubits & 3,706 & 986 & 8,700 & 2,744 \\
             Number of QEC rounds & 31,179 & 17,271 & 4,665 & 2,331 \\
             Total error probability & $8.1 \times 10^{-3}$ & $2.6 \times 10^{-3}$& $7.3 \times 10^{-3}$ & $4.1 \times 10^{-3}$ \\
        \end{tabular}
    \end{ruledtabular}
\end{table*}

\subsubsection{Cost of directly implementing Clifford and $T$ gates}
\label{sssec:cost-direct-impl-cliffords}

Using the estimates described in Section \ref{ssec:direct-impl-clifford}, we note that there are four logical patches to consider when estimating code distance. In terms of time requirements, CNOT and Hadamard gates require $3d + 4$ rounds, $S$/$S^\dagger$ gates require $3d/2 + 3$ rounds, $T$-like gates require up to $5d/2 + 4$ rounds, and $Z$ basis measurements require a single round. This brings our total number of rounds to $2318d + 3363$.

Using Eq.\ \ref{eq:distance-bound}, we find that for a physical error probability of $10^{-3}$, distance $d=12$ achieves a logical error probability of $3.9 \times 10^{-3}$, requiring 1,352 physical qubits for the patches and 31,179 QEC rounds. The factory in Table \ref{tab:t-factories} produces a $\ket{T}$ state with error probability $8.1 \times 10^{-6}$ on average once every 31.3 QEC rounds, meaning a single factory is sufficient. This factory uses 2,066 physical qubits, along with 288 physical qubits for storing $\ket{T}$ states. Combined with our 1,352 qubits for the logical circuit and routing, this leads to a total of 3,706 physical qubits. The additional logical qubit for storing $\ket{T}$ states increases the probability of a logical error to $4.9 \times 10^{-3}$, leading to a total error probability of $8.1 \times 10^{-3}$.

For a physical error probability of $10^{-4}$, an error probability of $6.9 \times 10^{-4}$ can be achieved with distance $d=6$, which requires 14,953 QEC rounds and 288 physical qubits. A $\ket{T}$ state needs to be generated every 29 rounds with an error probability of $1.3 \times 10^{-5}$. For this physical error probability, the factory in Table \ref{tab:t-factories} produces a $\ket{T}$ state on average every 18.05 rounds with error probability $4.7 \times 10^{-6}$. We use a single factory, which requires 522 physical qubits, along with 72 physical qubits for storing $\ket{T}$ states. Adding in our 392 qubits for the logical circuit and routing, this gives us a total of 986 physical qubits. The additional storage space for data qubits increases the probability of a logical error on qubits used in the quantum circuit to $8.6355 \times 10^{-4}$, leading to a total error probability of $2.6 \times 10^{-3}$.

\subsubsection{Cost of moving Clifford gates}
\label{sssec:cost-commute-cliffords}

If we choose to move Clifford gates through the circuit, we are left with a total of 386 Pauli $\pi/4$ rotations, each of which requires $d+1$ QEC rounds, and three joint Pauli measurements, which require $d$ QEC rounds each. Therefore our total number of QEC rounds is $389d + 386$. We also have six logical patches allocated for both the logical circuit and routing.

For a physical error probability of $10^{-3}$, distance $d = 11$ achieves a logical error probability of $2.8 \times 10^{-3}$, requiring 1,776 physical qubits and 4,665 QEC rounds. A $\ket{T}$ state needs to be produced once every 12 QEC rounds. We use the same $15$-to-$1$ factory as in Table \ref{tab:t-factories}, however a single factory is not sufficient for producing one $\ket{T}$ state every 12 rounds. Instead, we use three factories, which produce a single $\ket{T}$ state on average once every 10.4 QEC rounds and require 6,198 physical qubits for implementing the factories. The additional three logical qubits for storing $\ket{T}$ states increase the probability of a logical error on a qubit used in the quantum circuit to $4.2 \times 10^{-3}$, leading to a total error probability of $7.3 \times 10^{-3}$. The total number of physical qubits is 8,700.

For a physical error probability of $10^{-4}$, distance $d=5$ achieves a logical error probability of $1.4 \times 10^{-3}$ at a cost of 2,331 QEC rounds and 456 physical qubits. Using the same factory as in Table \ref{tab:t-factories} produces a $\ket{T}$ state with sufficiently low failure probability every 18.05 rounds, but a $\ket{T}$ is required every 6 rounds. Arranging four factories around the data qubits is sufficient to remove this bottleneck. These four factories require 2,088 physical qubits, and 200 physical qubits for storing $\ket{T}$ states. Adding this to our 456 physical qubits for the logical patches and routing leads to a total of 2,744 physical qubits. The extra four logical qubits for storing $\ket{T}$ states increase the probability of a logical error on a qubit used in the quantum circuit to $2.3 \times 10^{-3}$, which means the total error probability is $4.1 \times 10^{-3}$.

\subsection{Analysis}

As we can see from Fig.\ \ref{fig:results-summary}, there are still some significant overheads introduced from quantum error correction. The most optimistic error rates still require hundreds of physical qubits and thousands of QEC rounds, while at an error rate of $0.2\%$ this circuit requires tens of thousands of qubits and QEC rounds.

From the detailed costings of Section \ref{ssec:results}, we can identify several bottlenecks with these approaches. For physical qubits, the overhead mostly comes from $\ket{T}$ state factories: at a physical error rate $p=10^{-4}$ a single factory requires 522 physical qubits, nearly twice as many as required by the data qubits when implementing Clifford and $T$ gates directly. This is even more prominent when moving Clifford gates through the circuit, where of the 2,744 physical qubits required at error rate $10^{-4}$, 2,288 are for preparing and storing $\ket{T}$ states.  Although using fewer factories can reduce the number of physical qubits, this creates time bottleneck as the data qubits need to remain idle while $\ket{T}$ states are prepared.

Although it is expected that the overhead from such factories will become a less significant factor as we move towards larger quantum computations \cite{Litinski2019magicstate}, for early fault-tolerant quantum circuits these overheads are likely to be more costly. This could be improved via more efficient small footprint factories like the ones presented in \cite{Litinski2019magicstate}, as well as the use of error-mitigated $T$ gates \cite{Piveteau2021MitigatedT}.

When moving Clifford operations through the circuit, another space overhead comes from joint Pauli operations in the $Y$ basis. These require additional routing space and extra connectivity. Optimising the circuit to remove such measurements would also therefore reduce the routing overhead.

In time complexity, a significant bottleneck is the long sequences of Hadamard and $T$ gates which come from \texttt{gridsynth} decompositions. Of the 846 logical operations implemented in this circuit, 797 are either Hadamard or $T$ gates. The Hadamard gate is especially expensive, requiring $3d + 4$ QEC rounds. In practice, this means that more than half the QEC rounds are spent implementing Hadamard gates: at a physical error rate of $10^{-4}$, 9,042 of the 17,271 QEC rounds are spent implementing logical Hadamard gates. Time requirements can also be further reduced in general by using gate-based teleportation to execute gates in parallel, though this comes at a cost of more physical qubits \cite{Litinski2019}.

Finally it is worth emphasising that there are other ways in which these resource estimates can be improved above the quantum error correction layer, such as the use of different quantum algorithms \cite{Graves2023HydrogenMolecule} and decomposition techniques \cite{Kliuchnikov2022}.

\section{Conclusion}
\label{sec:conclusion}

As we enter the era of early fault-tolerant quantum computers, where quantum error correction is able to suppress errors on a logical qubit and basic logical gates are demonstrable, it is essential for us to understand the progress required for large-scale fault-tolerant quantum algorithms. Understanding the requirements of small applications is an important step in the process. In this work, we have analysed a minimal application: estimating the ground-state energy of the hydrogen molecule. We have used several techniques to reduce the estimated resources to approximately 900 physical qubits and 15,000 QEC rounds through implementing Clifford and $T$ operations directly, and approximately 2,700 physical qubits and 2,331 QEC rounds when implementing general Pauli $\pi/4$ rotations.

It is worth emphasising that even for this small application, the numbers of physical qubits and gates required is several orders of magnitude larger than what has been performed experimentally so far. There are a number of further optimisations which can be made across the quantum computing stack in the hope of reducing these estimates. At the algorithmic level, techniques such as qubitisation have been shown to produce asymptotically shorter quantum circuits \cite{Low2019hamiltonian, Low2017Qubitisation, Gilyen2019QSVT, Martyn2021Unification}, and could potentially offer improvements even for this minimal example \cite{Graves2023HydrogenMolecule}. Statistical phase estimation methods can allow reduced circuit depth in exchange for performing more samples \cite{Wan2022, Wang2022, Ding2022}, and are often stated as being particularly appropriate for the early fault-tolerant era for this reason. At the gate synthesis level, alternative techniques have produced circuits with a smaller $T$ count, at the cost of additional logical qubits \cite{Kliuchnikov2022}. When implementing $\pi/4$ joint Pauli rotations, the number of QEC rounds can be further reduced by implementing non-commuting rotations in parallel on separate patches before using teleportation to combine them, though this comes at a cost of more physical qubits \cite{Litinski2019}. Finally, improvements can be made to the implementation of non-Clifford gates which are more targeted towards early fault-tolerant quantum devices, such as the use of error mitigation when implementing faulty $T$ gates \cite{Piveteau2021MitigatedT}, avoiding the need for magic state distillation factories. Algorithms such as statistical phase estimation may remain well suited even in the presence of error mitigation \cite{Blunt2023Rigetti}.

A final note is that these estimates assume that quantum computers are affected specifically by depolarising noise \cite{Fowler2018, Litinski2019, Gidney2021}. While depolarising noise is easy to mathematically model, the physical noise that affects real-world devices is more complex and cannot necessarily be captured by such a model. An important direction of future work is investigating other more realistic noise models such as leakage and deriving similar scaling formulae to that presented in Eq.\ \ref{eq:logical-error-rate}.

\section*{Code Availability}

The source code for generating \& running the logical circuits, and estimating resources, is available on GitHub \cite{SourceCode}.

\begin{acknowledgments}
We thank Ophelia Crawford, Earl T. Campbell, Nicole Holzmann, Jacob M. Taylor and other Riverlane colleagues for insightful discussions, and Daniel Litinski for making his code for estimating the resource requirements of $\ket{T}$ factories publicly available.
\end{acknowledgments}

\bibliography{main}

\appendix

\section{Logical circuit figures}
\label{app:logical-circuit-figures}

\begin{figure*}
    \raggedright
    \begin{quantikz}
        \lstick{$\ket{0}$} & \gate{H} & \gate[2]{R(\theta_1)}\gateinput{$Z$} & \gate[2]{R(\theta_2)}\gateinput{$Z$} & \gate[2]{R(2\theta_1)}\gateinput{$Z$} & \gate[2]{R(\theta_2)}\gateinput{$Z$} & \gate[2]{R(2\theta_1)}\gateinput{$Z$} & \gate[2]{R(\theta_2)}\gateinput{$Z$} & \gate[2]{R(2\theta_1)}\gateinput{$Z$} & \gate[2]{R(\theta_2)}\gateinput{$Z$} & \gate[2]{R(\theta_1)}\gateinput{$Z$} & \qw \\
        \lstick{$\ket{\psi}$} & \gate{X} & \gateinput{$Z$} & \gateinput{$X$} & \gateinput{$Z$} & \gateinput{$X$} & \gateinput{$Z$} & \gateinput{$X$} & \gateinput{$Z$} & \gateinput{$X$} & \gateinput{$Z$} & \qw \\
        \lstick{$c_0$} & \cw & \cw & \cw & \cw & \cw & \cw & \cw & \cw & \cw & \cw & \cw \\
        \lstick{$c_1$} & \cw & \cw & \cw & \cw & \cw & \cw & \cw & \cw & \cw & \cw & \cw \\
        \lstick{$c_2$} & \cw & \cw & \cw & \cw & \cw & \cw & \cw & \cw & \cw & \cw & \cw
    \end{quantikz}

    \bigskip\bigskip
    
    \begin{quantikz}
        \qw & \gate{H} & \meter{} & \gate{X} & \gate{H} & \gate[2]{R(\theta_1)}\gateinput{$Z$} & \gate[2]{R(\theta_2)}\gateinput{$Z$} & \gate[2]{R(2\theta_1)}\gateinput{$Z$} & \gate[2]{R(\theta_2)}\gateinput{$Z$} & \gate[2]{R(\theta_1)}\gateinput{$Z$} & \gate{S^\dagger} & \gate{H} & \meter{} & \qw \\
        \qw & \qw & \qw & \qw & \qw & \gateinput{$Z$} & \gateinput{$X$} & \gateinput{$Z$} & \gateinput{$X$} & \gateinput{$Z$} & \qw & \qw & \qw & \qw \\
        \lstick{$c_0$} & \cw & \cwbend{-2} & \cwbend{-2} & \cw & \cw & \cw & \cw & \cw & \cw & \cwbend{-2} & \cw & \cw & \cw \\
        \lstick{$c_1$} & \cw & \cw & \cw & \cw & \cw & \cw & \cw & \cw & \cw & \cw & \cw & \cwbend{-3} & \cw \\
        \lstick{$c_2$} & \cw & \cw & \cw & \cw & \cw & \cw & \cw & \cw & \cw & \cw & \cw & \cw & \cw 
    \end{quantikz}

    \bigskip\bigskip
    
    \begin{quantikz}
        \qw & \gate{X} & \gate{H} & \gate[2]{R(\theta_1)}\gateinput{$Z$} & \gate[2]{R(\theta_2)}\gateinput{$Z$} & \gate[2]{R(\theta_1)}\gateinput{$Z$} & \gate{T^\dagger} & \gate{S^\dagger} & \gate{H} & \meter{} & \qw \\
        \lstick{$q_1$} & \qw & \qw & \gateinput{$Z$} & \gateinput{$X$} \qw & \gateinput{$Z$} & \qw & \qw & \qw & \qw & \qw \\
        \lstick{$c_0$} & \cw & \cw & \cw & \cw & \cw & \cwbend{-2} & \cw & \cw & \cw & \rstick{$\text{out}_0$} \cw \\
        \lstick{$c_1$} & \cwbend{-3} & \cw & \cw & \cw & \cw & \cw & \cwbend{-3} & \cw & \cw & \rstick{$\text{out}_1$} \cw \\
        \lstick{$c_2$} & \cw & \cw & \cw & \cw & \cw & \cw & \cw & \cw & \cwbend{-4} & \rstick{$\text{out}_2$} \cw
    \end{quantikz}
    \caption{Logical quantum circuit for iterative QPE to three bits of precision.}
    \label{fig:logical-iterative-qpe}
\end{figure*}
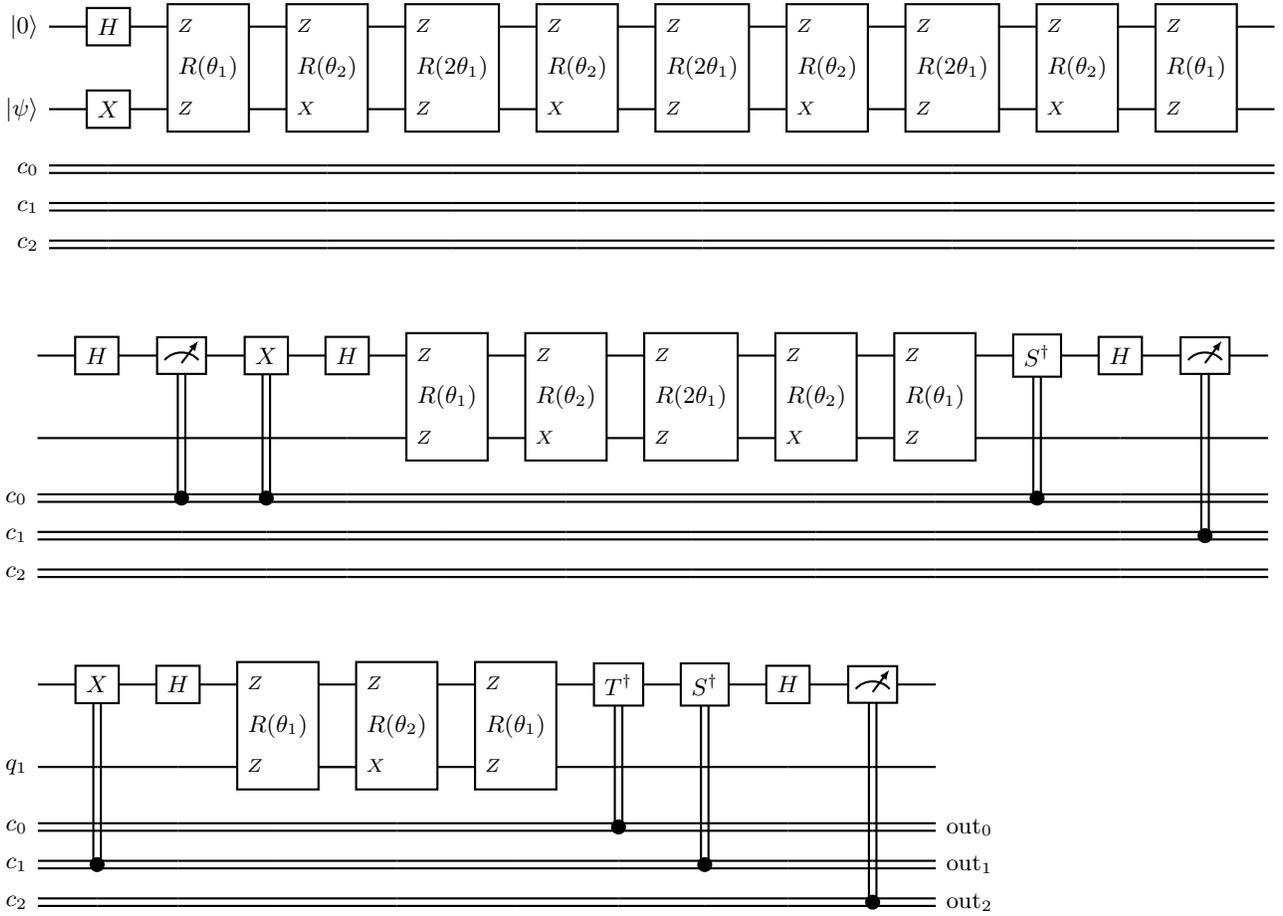

The complete logical circuits for estimating the ground-state energy of the hydrogen molecule using textbook QPE and iterative QPE are given in Figures \ref{fig:logical-textbook-qpe} and \ref{fig:logical-iterative-qpe}, respectively. In the iterative QPE circuit, the classically-controlled $X$ gates are used to reset the measured qubits to the $\ket{0}$ state.

\begin{figure*}
    \raggedright
    \begin{quantikz}
        \lstick{$\ket{0}$} & \gate{H} & \qw & \qw & \qw & \qw & \qw & \qw & \qw & \qw & \gate[4]{R(\theta_1)}\gateinput{$Z$} & \qw \\
        \lstick{$\ket{0}$} & \gate{H} & \qw & \qw & \qw  & \gate[3]{R(\theta_1)}\gateinput{$Z$} & \gate[3]{R(\theta_2)}\gateinput{$Z$} & \gate[3]{R(2\theta_1)}\gateinput{$Z$} & \gate[3]{R(\theta_2)}\gateinput{$Z$} & \gate[3]{R(\theta_1)}\gateinput{$Z$} & \qw & \qw \\
        \lstick{$\ket{0}$} & \gate{H} & \gate[2]{R(\theta_1)}\gateinput{$Z$} & \gate[2]{R(\theta_2)}\gateinput{$Z$} & \gate[2]{R(\theta_1)}\gateinput{$Z$} & \qw & \qw & \qw & \qw & \qw & \qw & \qw \\
        \lstick{$\ket{\psi}$} & \gate{X} & \gateinput{$Z$} & \gateinput{$X$} & \gateinput{$Z$} & \gateinput{$Z$} & \gateinput{$X$} & \gateinput{$Z$} & \gateinput{$X$} & \gateinput{$Z$} & \gateinput{$Z$} & \qw
    \end{quantikz}

    \bigskip\bigskip
    
    \begin{quantikz}
        \qw & \gate[4]{R(\theta_2)}\gateinput{$Z$} & \gate[4]{R(2\theta_1)}\gateinput{$Z$} & \gate[4]{R(\theta_2)}\gateinput{$Z$} & \gate[4]{R(2\theta_1)}\gateinput{$Z$} & \gate[4]{R(\theta_2)}\gateinput{$Z$} & \gate[4]{R(2\theta_1)}\gateinput{$Z$} & \gate[4]{R(\theta_2)}\gateinput{$Z$} & \gate[4]{R(\theta_1)}\gateinput{$Z$} & \gate{H} & \phase{-\pi/2} & \qw\\
        \qw & \qw & \qw & \qw & \qw & \qw & \qw & \qw & \qw & \qw & \ctrl{-1} & \qw \\
        \qw & \qw & \qw & \qw & \qw & \qw & \qw & \qw & \qw & \qw & \qw & \qw \\
        \qw & \gateinput{$X$} & \gateinput{$Z$} & \gateinput{$X$} & \gateinput{$Z$} & \gateinput{$X$} & \gateinput{$Z$} & \gateinput{$X$} & \gateinput{$Z$} & \qw & \qw & \qw
    \end{quantikz}

    \bigskip\bigskip
    
    \begin{quantikz}
        \qw & \qw &\phase{-\pi/4} & \qw & \qw & \meter{} & \rstick{$\text{out}_0$} \cw \\
        \qw & \gate{H} & \qw & \phase{-\pi/2} & \qw & \meter{} & \rstick{$\text{out}_1$} \cw \\
        \qw & \qw & \ctrl{-2} & \ctrl{-1} & \gate{H} & \meter{} & \rstick{$\text{out}_2$} \cw \\
        \qw & \qw & \qw & \qw & \qw & \qw & \qw
    \end{quantikz}
    \caption{Logical quantum circuit for textbook QPE to three bits of precision.}
    \label{fig:logical-textbook-qpe}
\end{figure*}
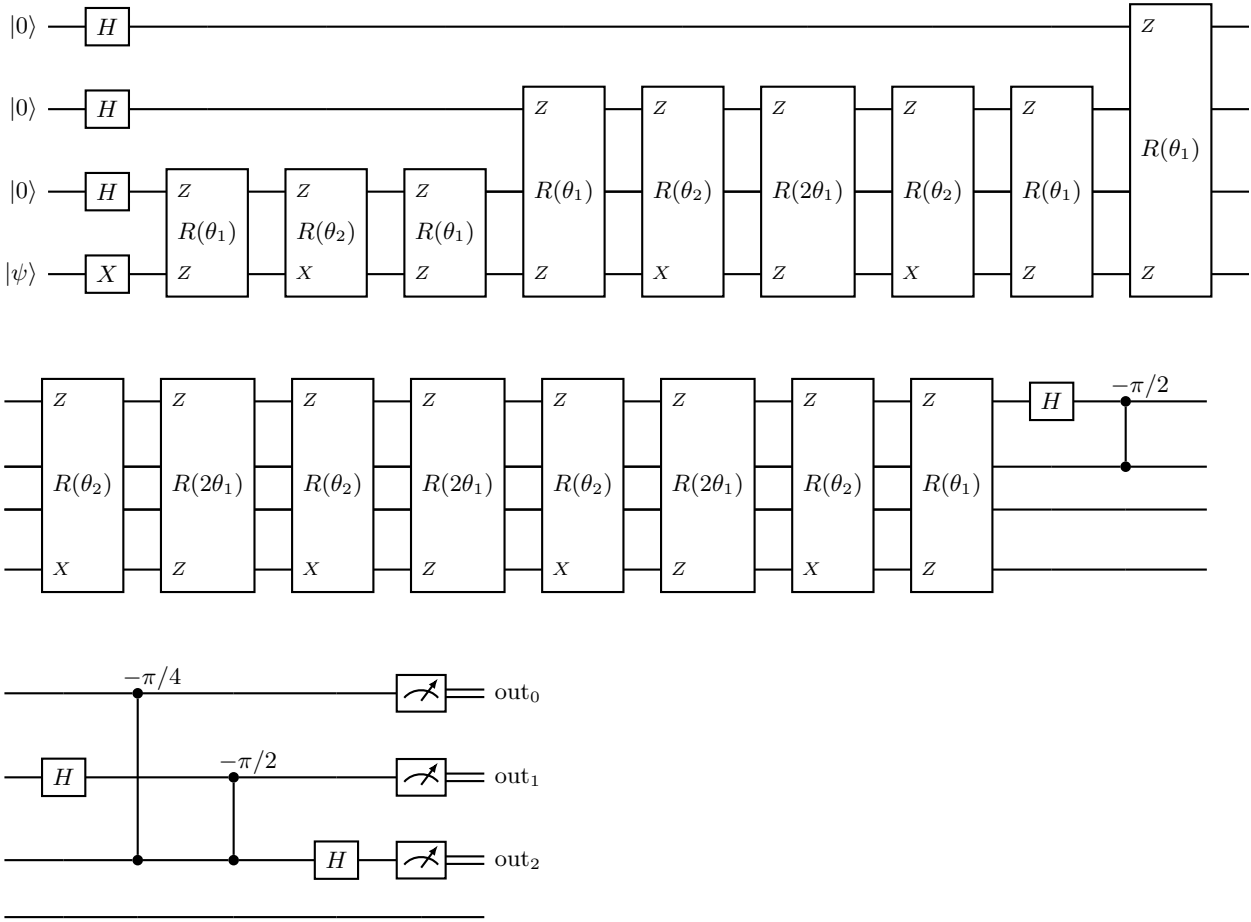

\section{Implementation of $T$-like gates}
\label{app:implementing-t-like-gates}

\begin{figure*}
    \begin{subfigure}[b]{0.3\linewidth}
        \begin{quantikz}
            & \gate[3, nwires={2}]{M}\gateinput{$Z$} & \gate{Z \text{ or } S} & \gate{Z} & \qw \\
            & & \cwbend{-1} & & \\
            \lstick{\ket{T}} & \gateinput{$Z$} & \qw & \gate{M_X} \vcw{-2} &
        \end{quantikz}
        \caption{$T$ gate followed by an $S$ gate}
        \label{fig:ts-gate-circuit}
    \end{subfigure}
    \begin{subfigure}[b]{0.3\linewidth}
        \begin{quantikz}
            & \gate[3, nwires={2}]{M}\gateinput{$Z$} & \gate{S} & \gate{Z} & \qw \\
            & & \cwbend{-1} & & \\
            \lstick{\ket{T}} & \gateinput{$Z$} & \gate{Z} & \gate{M_X} \vcw{-2} &
        \end{quantikz}
        \caption{$T$ gate followed by a $Z$ gate}
        \label{fig:tz-gate-circuit}
    \end{subfigure}
    \begin{subfigure}[b]{0.3\linewidth}
        \begin{quantikz}
            & \gate[3, nwires={2}]{M}\gateinput{$Z$} & \gate{Z \text{ or } S} & \gate{Z} & \qw \\
            & & \cwbend{-1} & & \\
            \lstick{\ket{T}} & \gateinput{$Z$} & \gate{Z} & \gate{M_X} \vcw{-2} &
        \end{quantikz}
        \caption{$T$ gate followed by an $S$ gate and a $Z$ gate}
        \label{fig:tsz-gate-circuit}
    \end{subfigure}
    \caption{Example circuits for implementing $T$-like gates through joint $Z \otimes Z$ measurements. In (\ref{fig:ts-gate-circuit}), a $T$ gate followed by an $S$ gate is implemented with the same cost as computing a single $T$ gate as shown in Figure \ref{fig:t-gate-circuit}, by adjusting the Clifford operations performed after the measurement. Note that in (\ref{fig:ts-gate-circuit}), either a $Z$ gate is implemented or an $S$ gate is implemented, depending on the outcome of the joint $Z \otimes Z$ measurement. In (\ref{fig:tz-gate-circuit}) and (\ref{fig:tsz-gate-circuit}), we use the same techniques for implementing a $T$ gate followed by a $Z$ gate, and a $T$ gate followed by an $S$ gate and a $Z$ gate, respectively. All circuits require an additional qubit initialised in the $\ket{T}=(\ket{0} + e^{i\pi/2}\ket{1})/\sqrt{2}$ state.}
    \label{fig:t-like-gate-circuits}
\end{figure*}
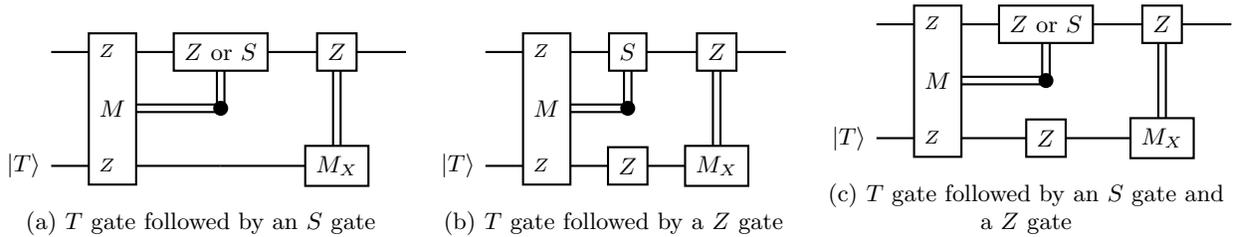

Here we explain how to implement any sequence of $T$/$T^\dagger$, $S$/$S^\dagger$ and $Z$ gates. This is done in three steps: first, by removing inverse gates by noting that $S^\dagger = ZS$ and $T^\dagger = ZST$; second, by combining the different phase gates into a sequence consisting of at most one $Z$ gate, one $S$ gate, and one $T$ gate; and third, by absorbing the non-$T$ gates into the conditional operations in Figure \ref{fig:t-gate-circuit}. In Figure \ref{fig:ts-gate-circuit}, we give an example of implementing a $T$ gate followed by an $S$ gate using this procedure: we combine the $S$ gate following the $T$ gate with the conditional $S$ gate, to create a gate where either a $Z$ operation is applied at no extra cost, or an $S$ gate is applied, depending on the joint $Z \otimes Z$ measurement outcome. Similar circuits can also be generated for $TZ$ and $TSZ$ gate sequences, as shown in Figures \ref{fig:tz-gate-circuit} and \ref{fig:tsz-gate-circuit}, respectively. Note that the $Z$ operation on the auxiliary qubit in Figures \ref{fig:tz-gate-circuit} and \ref{fig:tsz-gate-circuit} can be implemented at no extra cost by simply inverting the result of the $X$ basis measurement.

\section{Translation rules for Pauli rotations}
\label{app:commutation-rules}

\begin{figure*}
    \begin{subfigure}[b]{\linewidth}
        \begin{quantikz}
            & \gate{R_P(\pi/2)} \qwbundle[alternate]{} & \gate{R_{P'}(\pi/4)} \qwbundle[alternate]{} & \qwbundle[alternate]{}
        \end{quantikz}
        \hspace{0.3cm}=\hspace{0.1cm}
        \begin{quantikz}
            & \gate{R_{P'}(\pi/4)} \qwbundle[alternate]{} & \gate{R_{P}(\pi/2)} \qwbundle[alternate]{} & \qwbundle[alternate]{}
        \end{quantikz}
        \caption{Commuting $P$ and $P'$}
        \label{fig:t-gate-commuting}
    \end{subfigure}

    \vspace{0.6cm}

    \begin{subfigure}[b]{\linewidth}
        \begin{quantikz}
            & \gate{R_P(\pi/2)} \qwbundle[alternate]{} & \gate{R_{P'}(\pi/4)} \qwbundle[alternate]{} & \qwbundle[alternate]{}
        \end{quantikz}
        \hspace{0.3cm}=\hspace{0.1cm}
        \begin{quantikz}
            & \gate{R_{iPP'}(\pi/4)} \qwbundle[alternate]{} & \gate{R_{P}(\pi/2)} \qwbundle[alternate]{} & \qwbundle[alternate]{}
        \end{quantikz}
        \caption{Non-commuting $P$ and $P'$}
        \label{fig:t-gate-noncommuting}
    \end{subfigure}

    \vspace{0.6cm}

    \begin{subfigure}[b]{\linewidth}
        \begin{quantikz}
            & \gate{R_P(\pi/2)} \qwbundle[alternate]{} & \gate{M_{P'}} \qwbundle[alternate]{} & \qwbundle[alternate]{}
        \end{quantikz}
        \hspace{0.3cm}=\hspace{0.1cm}
        \begin{quantikz}
            & \gate{M_{P'}} \qwbundle[alternate]{} & \gate{R_{P}(\pi/2)} \qwbundle[alternate]{} & \qwbundle[alternate]{}
        \end{quantikz}
        \caption{Commuting $P$ and $P'$}
        \label{fig:measurement-commuting}
    \end{subfigure}

    \vspace{0.6cm}

    \begin{subfigure}[b]{\linewidth}
        \begin{quantikz}
            & \gate{R_P(\pi/2)} \qwbundle[alternate]{} & \gate{M_{P'}} \qwbundle[alternate]{} & \qwbundle[alternate]{}
        \end{quantikz}
        \hspace{0.3cm}=\hspace{0.1cm}
        \begin{quantikz}
            & \gate{M_{iPP'}} \qwbundle[alternate]{} & \gate{R_{P}(\pi/2)} \qwbundle[alternate]{} & \qwbundle[alternate]{}
        \end{quantikz}
        \caption{Non-commuting $P$ and $P'$}
        \label{fig:measurement-noncommuting}
    \end{subfigure}
    \caption{Moving $n$-qubit $\pi/2$ Pauli rotations past other Pauli operations. (\ref{fig:t-gate-commuting}) A $\pi/2$ rotation in Pauli basis $P$ can be swapped with a $\pi/4$ rotation in a commuting Pauli basis $P'$. (\ref{fig:t-gate-noncommuting}) A $\pi/2$ rotation in a Pauli basis $P$ can also be swapped with a $\pi/4$ rotation in a non-commuting basis $P'$, by modifying the basis of the $\pi/4$ rotation to $iPP'$. These rules also apply to moving $\pi/2$ rotations past Pauli measurements, as shown in (\ref{fig:measurement-commuting}) and (\ref{fig:measurement-noncommuting}).}
    \label{fig:commuting-pauli-rotations}
\end{figure*}
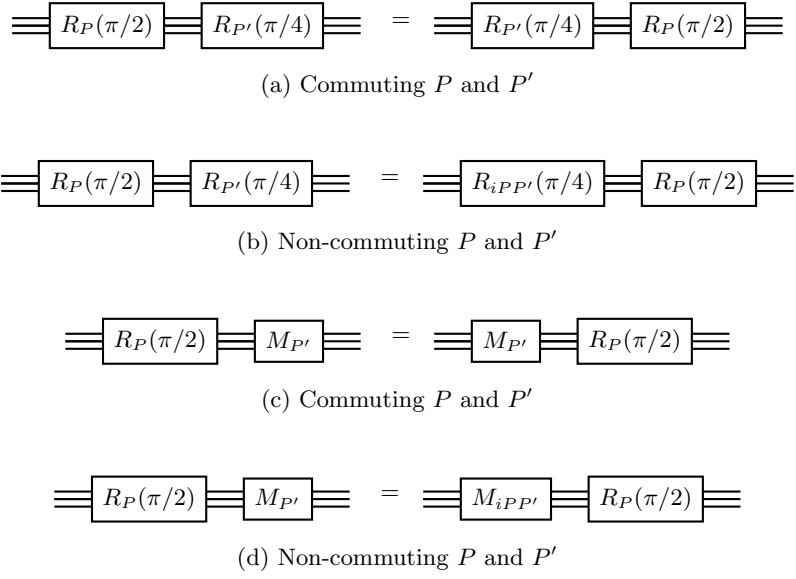

Here we explain the rules for manipulating Pauli rotations used in Section \ref{ssec:commute-clifford}. In general, a $\pi/2$ rotation in some Pauli basis $P$ can be moved past a $\pi/4$ operation in some Pauli basis $P'$ without modification to either basis if $P$ and $P'$ commute, or by modifying $P'$ to $iPP'$ if they do not commute. These rules are presented graphically in Figure \ref{fig:commuting-pauli-rotations} \footnote{Note that our rules for moving Pauli rotations past measurements are slightly different from those presented in Ref.~\cite{Litinski2019}. This is because unlike the circuits in Ref.~\cite{Litinski2019} where there is a single layer of measurement gates at the end of the circuit, the iterative QPE circuit in Figure \ref{fig:iqpe_circuit} features mid-circuit measurements and therefore other computations happen after a measurement gate.}. These rules can also be used to move $\pi$ rotations as well, by noting that $R_P(\pi) = R_P(\pi/2) \cdot R_P(\pi/2)$.

\begin{figure}
    \begin{quantikz}
        & \gate[2]{R(\frac{\pi}{2})}\gateinput{$Z$} & \qw & \qw \\
        & \gateinput{$X$} & \gate{R_Z(\frac{\pi}{4})} & \qw
    \end{quantikz}
    \hspace{0.2cm}=
    \begin{quantikz}
        & \gate[2]{R(\frac{\pi}{4})}\gateinput{$Z$} & \gate[2]{R(\frac{\pi}{2})}\gateinput{$Z$} & \qw \\
        & \gateinput{$Y$} & \gateinput{$X$} & \qw
    \end{quantikz}
    \caption{Moving a two-qubit $\pi/2$ $Z \otimes X$ rotation past a single-qubit $\pi/4$ $Z$ rotation leads to a two-qubit $\pi/4$ $Z \otimes Y$ rotation.}
    \label{fig:commuting-multi-qubit-rotations}
\end{figure}
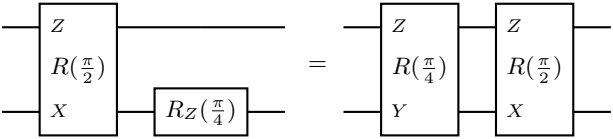

It is important however to note that these remaining non-Clifford operations can be more complicated than the single-qubit non-Clifford operations in the original logical circuit. An example of this is shown in Figure \ref{fig:commuting-multi-qubit-rotations}, where a $\pi/2$ $Z \otimes X$ rotation, such as the one that shows up in the CNOT circuit of Figure \ref{fig:cnot-pauli-rotation}, is moved past a $\pi/4$ $Z$ rotation on the second qubit. Noting that $Z \otimes X$ does not commute with $I \otimes Z$, the basis for the $\pi/4$ rotation becomes

$$i (Z \cdot I) \otimes (X \cdot Z) = i Z \otimes (-i Y) = Z \otimes Y,$$

\noindent
thus what was originally a $\pi/4$ rotation across a single qubit has now become a $\pi/4$ rotation across multiple qubits.

\section{Arrangement of $\ket{T}$ state factories}
\label{app:factory-arrangements}

\begin{figure*}
    \begin{subfigure}[b]{\linewidth}
        \includegraphics[width=0.5\linewidth]{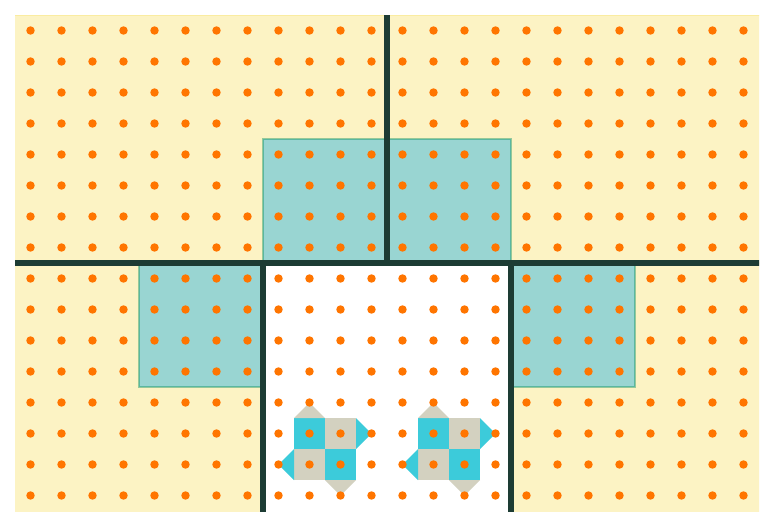}
        \caption{Directly implementing Clifford and $T$ gates}
        \label{fig:factory-placement-do-cliffords}
    \end{subfigure}
    \begin{subfigure}[b]{\linewidth}
        \includegraphics[width=0.5\linewidth]{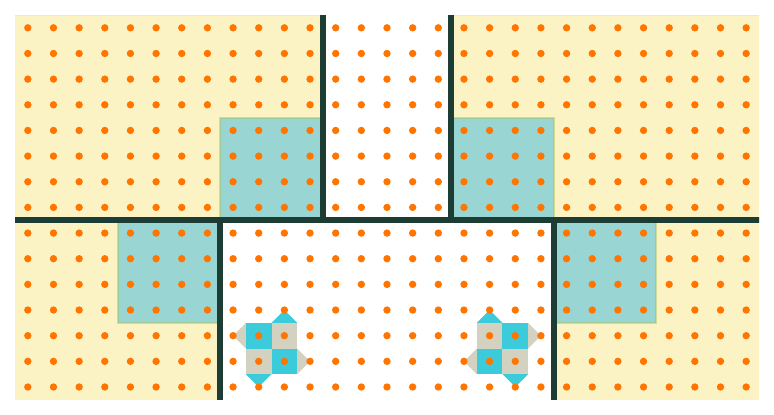}
        \caption{Moving Clifford gates}
        \label{fig:factory-placement-pauli-rotations}
    \end{subfigure}
    \caption{Arrangements of $\ket{T}$ state factories around the routing spaces for (\ref{fig:factory-placement-do-cliffords}) implementing Clifford and $T$ operations directly and (\ref{fig:factory-placement-pauli-rotations}) commuting Clifford operations. Green space denotes storage space for $\ket{T}$ states produced by the factories, which are denoted in yellow. Note that the full factories are not shown due to size.}
    \label{fig:factory-arrangements}
\end{figure*}

In Figure \ref{fig:factory-arrangements} we show how to arrange magic state factories around the logical patches and routing space for both implementing Clifford and $T$ gates directly and moving Clifford gates through the circuit. In both cases, it is easily possible to arrange up to four factories around the logical patches. It is also possible to arrange even more factories, however this might come at the cost of additional routing space. For simplicity we stick with up to four factories.

Note that in Figure \ref{fig:factory-placement-pauli-rotations} there is some unused space at the top of the arrangement. This space has been left empty for ease of symmetry with the arrangement, but could be used as additional factory space. Such unused qubits are not included as part of the resource estimates presented in Section \ref{sec:qec-overheads}.

We also stress that the space in yellow is not the full space required for the factories, but simply an indicator of what space the factories can be placed in. The green lines denoting the boundaries between the factories and routing space should be seen as extending beyond the limits of Figure \ref{fig:factory-arrangements}, to however much space is required for individual factories.

\end{document}